%% file: paper.tex
\documentclass[11pt,a4paper,DIV11]{scrartcl}
\pdfoutput=1
\usepackage[utf8]{inputenc}
\usepackage[T1]{fontenc}
\usepackage{lmodern}
\usepackage[british]{babel}
\usepackage{amsmath}
\usepackage{amssymb}
\usepackage{amsfonts}
\usepackage{color}
\usepackage{float}
\usepackage{stmaryrd}
\usepackage{url}
\usepackage{hyperref}
\usepackage{graphicx,psfrag}
\usepackage{placeins,bbm}
\usepackage{subfigure}
\usepackage{authblk}
\usepackage{booktabs,adjustbox} 
\DeclareOldFontCommand{\bf}{\normalfont\bfseries}{\mathbf}

\newcommand{\Nf}{N_f}
\newcommand{\Nff}{N_f^{(F)}}
\newcommand{\Nfa}{N_f^{(A)}}
\newcommand{\wt}{\tilde{w}}
\newcommand{\kf}{\ensuremath{\kappa_F}}
\newcommand{\ka}{\ensuremath{\kappa_A}}
\newcommand{\su}[1]{\ensuremath{\text{SU}(#1)}}

\title{%
	Lattice simulations of a gauge theory with mixed adjoint-fundamental matter}
\author[1,2]{Georg Bergner\thanks{georg.bergner@uni-jena.de}}
\author[3]{Stefano Piemonte\thanks{stefano.piemonte@ur.de}}
\affil[1]{University of Jena, Institute for Theoretical Physics,\\ 
	Max-Wien-Platz 1, D-07743 Jena, Germany}
\affil[2]{University of M\"unster, Institute for Theoretical Physics, 
	Wilhelm-Klemm-Str.~9, D-48149 M\"unster, Germany}
\affil[3]{University of Regensburg, Institute for Theoretical Physics, 
	Universit\"atsstr.~31, D-93040 Regensburg, Germany}
\date{\today}
\begin{document}
\maketitle
\begin{abstract}
In this article we summarize our efforts in simulating Yang-Mills theories coupled to matter fields transforming under the fundamental and adjoint representations of the gauge group. In the context of composite Higgs scenarios, gauge theories with mixed representation fields have been suggested to describe the fundamental interactions well beyond the electroweak unification scale, and they are also closely related to supersymmetric QCD. In addition, they are studied as deformations of theories with pure adjoint matter in the context of adiabatic continuity. We provide some first results for bare parameter tuning and interdependence of the two representations. We also investigate how the chiral symmetry breaking or a conformal scenario can be realized and checked in such theories.
\end{abstract}
\section{Gauge theories with adjoint and fundamental fermions}
In the last two decades, there has been a substantial effort to extend lattice Monte-Carlo simulations from quantum chromodynamics (QCD) toward the full landscape of gauge theories including different numbers of fermion fields transforming in the fundamental representation of the gauge group. Higher fermion representations have been also considered, most notably the adjoint representation of SU(2) and SU(3), and the sextet representation of SU(3). The motivations for these studies have been the search for an extension of the Standard Model or the consideration of supersymmetric gauge theories. The first studies of gauge theories coupled to fermions in two different representations have been published very recently \cite{DeGrand:2016pgq,Ayyar:2017qdf,Cossu:2019hse}. Such a setup enhances substantially the possibilities for model building and investigations of general theoretical questions. Our study of a mixed representation setup with an SU(2) gauge theory coupled to two Dirac fermions in the fundamental and one Majorana fermion in the adjoint representation has several motivations.

The first aim is an exploratory study towards the investigations of supersymmetric QCD (SQCD). SQCD is described by SU($N_c$) gauge fields coupled to fermionic gluinos in the adjoint representation of the gauge group as well as $\Nf$ fermionic quark fields and scalar squark fields in the fundamental representation. Depending on $N_c$ and $\Nf$, different phases of the theory are expected. In particular there are predictions for the appearance of an IR conformal fixed point in this parameter space. There have been several attempts towards a simulation of SQCD on the lattice \cite{Giedt:2009yd,Costa:2017rht,Wellegehausen:2018opt,Bergner:2018znw}, but so far no real and complete investigation has been possible. Since there has not been much experience with numerical simulations of mixed representations, it is important to understand first the theory without scalar fields, as a first step toward SQCD.

A second aim is related to possible composite Higgs theories. Theories with fermions in higher representations have been already investigated in this context. One of the most famous examples is minimal walking technicolor (MWT), an SU(2) gauge theory with $\Nf=2$ Dirac fermions in the adjoint representation. As shown by numerical investigations, the mass anomalous dimension at the IR fixed point of this theory is quite small, which makes it less favorable for a Standard Model extension \cite{DelDebbio:2010hx,DeGrand:2011qd,Patella:2012da}. Theories with smaller $\Nf$ lead to a larger value of the mass anomalous dimension \cite{Athenodorou:2014eua,Bergner:2017gzw}, but they don't provide the right particle content for the coupling to the Standard Model. One possible way out is to combine different representations. The matter content for a coupling of the SU(2) gauge theory to the Standard Model are two fundamental fermions and the theory is driven towards the conformal or walking limit by additional adjoint fermion flavors. The adjoint matter is not charged under the gauge groups of the Standard Model. This minimizes the matter content of a possible Standard Model extension  compared to an approach with only fundamental matter since the adjoint representation requires a smaller number of flavors to reach near conformality.

The theory with two fundamental and one adjoint Dirac flavor, the so-called ultra minimal walking technicolor (UMWT), has been suggested as a strongly interacting completion of the Standard Model \cite{Ryttov:2008xe}. Due to the relation to SQCD, we are considering here only one Majorana fermion in the adjoint representation. This might nevertheless already be sufficient for near conformality since the analysis of the theory has so far only been based on perturbative estimates. The theory can, furthermore, be considered as a deformation SU(2) gauge theory with two fermions in the fundamental representation, which has been considered in several investigations as a composite Higgs model, see for example numerical studies in \cite{Lewis:2011zb,Hietanen:2014xca,Arthur:2016dir,Drach:2017btk}. 

Another relation of our investigations in the context of composite Higgs is the approach of a fine grained control of the running of the gauge coupling by different mass scales. This has been suggested and investigated for theories with a large number of flavours in the fundamental representation of SU(3) \cite{Brower:2015owo,Hasenfratz:2016gut}. The fermion fields have been split in a set of $(\Nf)_h$ heavy and  $(\Nf)_l$ light fermions. However, this appears more natural in the context of mixed representations as there is no symmetry suggesting an equality of the masses of the different representations.
Our studies might also provide additional insights for the general investigations 
with multiple fermion representations, which appear in the context of composite Higgs theories and partial compositeness \cite{Ferretti:2014qta}. There are so far only a limited number of lattice studies in this context \cite{DeGrand:2016pgq,Ayyar:2017qdf,Cossu:2019hse}.

The third line of motivation is related to a predicted analytic continuity between confinement of strongly coupled gauge theories and confinement in a semiclassical small circle regime. This provides a better analytic control in investigations of the relevance of non-perturbative semiclassical contributions in the confinement mechanism. The phenomenon is well understood in supersymmetric Yang-Mills theory (SYM) and also confirmed by numerical simulations \cite{Bergner:2014dua,Bergner:2018unx}. Theoretical studies have been based on the assumption that this can be extended towards gauge theories with a larger number of adjoint fermion flavors than the one Majorana fermion corresponding to SYM. It is difficult to verify these findings since already at one Dirac flavor, the theory becomes nearly conformal. 
The extension of SYM by fermion fields in the fundamental representation enlarges the space of possible applications for analytic continuity \cite{Cherman:2016hcd,Kanazawa:2019tnf}. It might also help to relate the SYM confinement with the confinement of QCD in a continuous way.

The current study represents the essential first step for all of these investigations, being an exploration of the parameter space spanned by the two mass parameters of the different representations and the gauge coupling. The study of the scaling of the meson mass spectrum close to the chiral limit provides a clear picture helping to distinguish the signals of a chiral symmetry breaking scenario from a conformal theory. In particular, the main target of the present investigations is the deformation of the spectrum of lowest mesonic states induced by the addition of fermions in a different representation. As a first step, we aim to identify possible unphysical bulk phases, which appear for higher representations and in particular in the context of near conformal theories. An important cross-check of our current first studies is also the connection to pure adjoint and fundamental limits.

\section{A theory with two different fermion representations on the lattice}

The first step for Monte-Carlo simulations is the lattice discretization of the continuum action. In our numerical simulations the gauge part of the lattice action is represented by the Wilson gauge action built from plaquettes $U_p$ of link variables $U$ in fundamental representation of SU($N_c$). The fermionic part comprises a Dirac-Wilson clover improved action for $\Nff$ fermions in the fundamental and $\Nfa$ fermions in the adjoint representation. 
In its basic form the complete lattice action reads
\begin{equation}
\mathcal{S}_L =
\beta \sum_p\left(1-\frac{1}{N_c}\mbox{Re}\,\mathrm{tr} U_p\right)
+\sum_{x,y}\sum_{n_f=1}^{\Nff} \bar{\psi}_x^{n_f}(D^{(F)}_w)_{xy}\psi_y^{n_f}+\sum_{x,y}\sum_{n_f=1}^{\Nfa} \bar{\psi}_x^{n_f}(D^{(A)}_w)_{xy}\psi_y^{n_f}\, ,
\end{equation}
where $D_w^{(F)}$ ($D_w^{(A)}$) is the clover Wilson-Dirac operator in the fundamental (adjoint) representation.
These operators depend on the hopping parameter $\kf$ ($\ka$), which is related to the bare fermion mass $m_0$ in the respective representation via
$\kappa=1/(2m_0+8)$. Like in our previous studies, the link fields in $D_w^{(A)}$ are converted to the adjoint representation. The two clover parameters have been tuned by a one-loop perturbative calculation \cite{Musberg:2013foa}, that has already provided significant improvements in our previous studies of pure SYM.

Our simulation program allows flexible simulation for an arbitrary number of fermions in the adjoint and fundamental representation of an $\mathrm{SU}(N_c)$ gauge group.
In the current work we consider one adjoint Majorana fermion, effectively $ \Nfa=\frac{1}{2}$, simulated with the rational Hybrid Monte-Carlo 
algorithm (RHMC), and two degenerate flavors in the fundamental representation ($\Nff=2$) which are represented by the standard Hybrid Monte-Carlo algorithm (HMC).

\section{Phase transitions at strong coupling}
\begin{figure}
    \subfigure[Pure adjoint theory\label{bulk_adj}]
	{\includegraphics[width=0.47\textwidth]{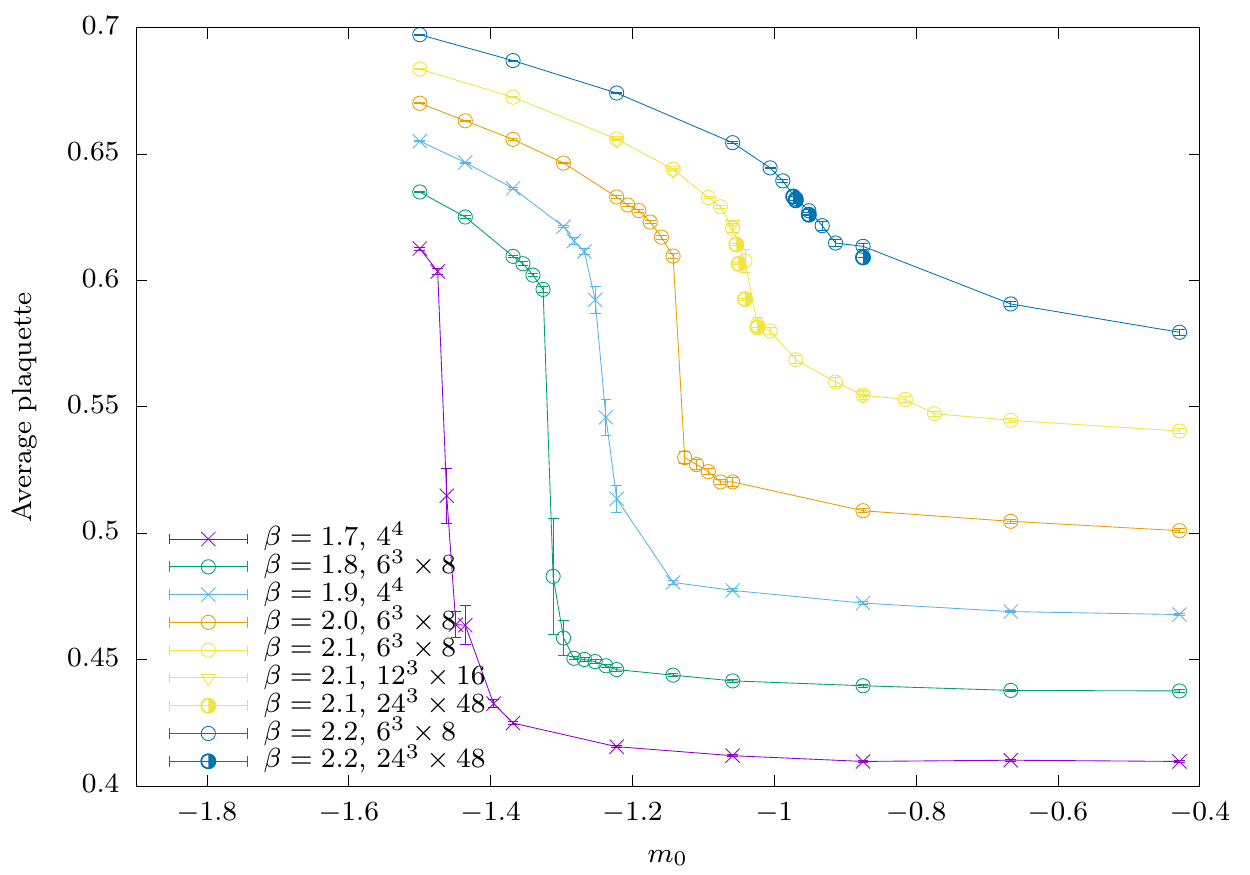}}
	\subfigure[$\kf$ dependence\label{bulk_kf}]
	{\includegraphics[width=0.47\textwidth]{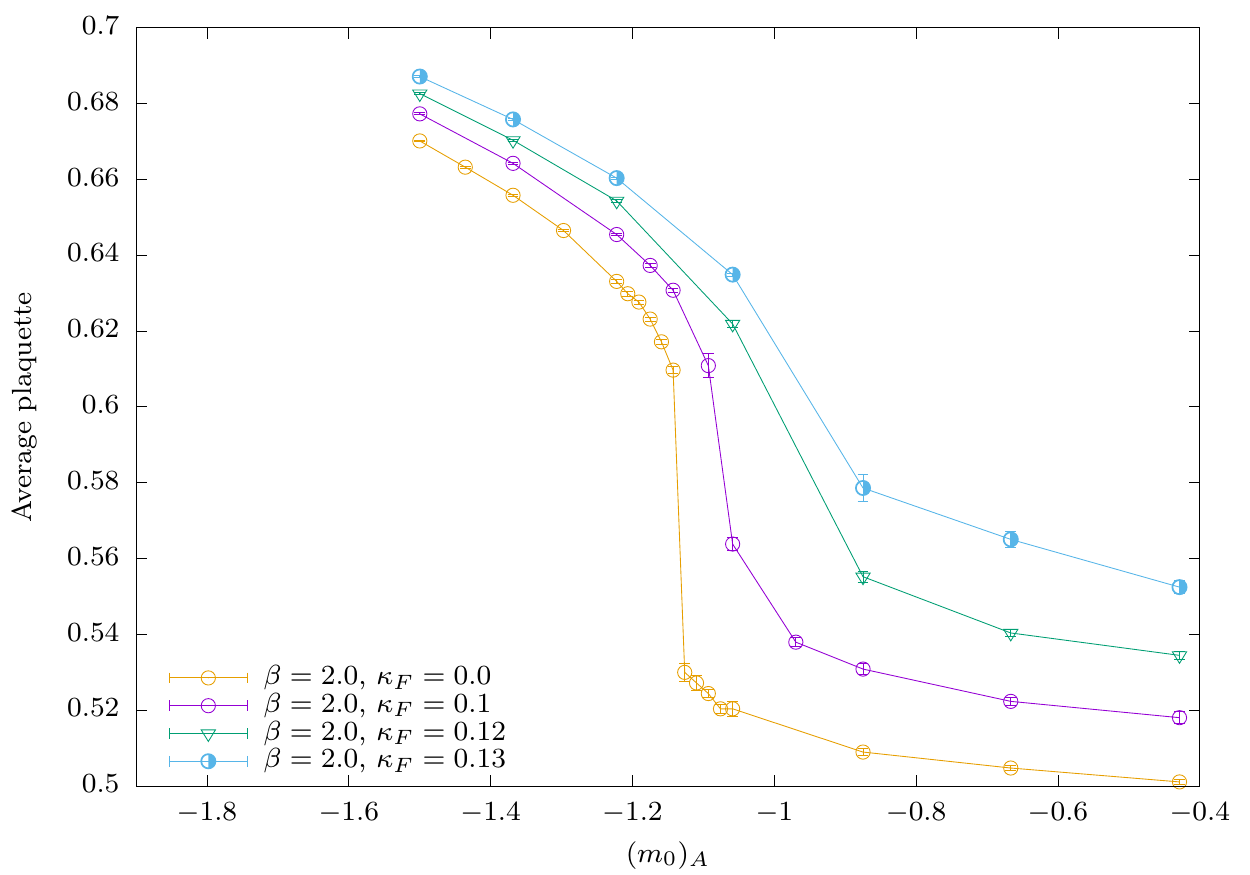}}
	\caption{The strong coupling phase transitions shown by discontinuity of the average plaquette as a function of the bare mass parameter. Figure \ref{bulk_adj} shows the pure adjoint theory ($\kf=0$). The transition can be observed for different lattice sizes of $N_s\times N_t=4^4,\quad 6^3\times 8,\quad 12^3\times 16,\quad 24^3\times 48$. The dependence on mass of the fundamental fermions is shown in Figure \ref{bulk_kf}. \label{fig:bulk} }
\end{figure}
In a first study of the theory we map out the phase diagram on small lattices to identify possible unphysical bulk phases appearing in the context of IR conformal theories. In the bare parameter space of these theories, the strong coupling confining regime has to be separated from the conformal phase, which corresponds to the range of gauge couplings attracted by the IR fixed point. Such kind of behaviour has been documented for example for MWT or theories with a large number of fundamental flavours. Since we don't know a priori how far our theory from the conformal window, we have to consider the possibility of bulk transitions.

Bulk transitions have to be considered in a more general context. Pure SU(2) Yang-Mills theory has a cross-over from weak to strong couplings which becomes a bulk transition when an additional adjoint Wilson gauge action is coupled to the theory. Therefore it is natural to expect bulk transitions for any theory with fermions in the adjoint representation. Moreover, there are a number of evidences for the bulk phases for theories with fermions in higher representation of the gauge group.

As first investigation, we monitor the expectation value of plaquette on small lattices as a function of $\beta$ and $\kappa$. Similar studies have been done in earlier investigations in order to identify the bulk phase of MWT \cite{Catterall:2008qk} and SU(2) gauge theory with one adjoint Dirac flavor \cite{Athenodorou:2014eua}. In the present case, the transition has to be mapped out in the space of $\beta$, $\kf$, and $\ka$. A scan of the parameter space is shown in Fig.~\ref{fig:bulk}. We have observed that the strongest transition appears in the pure adjoint case ($\kf=0$). The discontinuity gets weaker when the fundamental fermion part is added, see Fig.~\ref{bulk_kf}. In the present study, we want to connect our investigations to the pure adjoint and pure fundamental limit. Therefore we have limited our main studies to the range of beta values above the transition in the pure adjoint case, $\beta\geq 2.1$. Despite the fact that only one Majorana fermion has been considered here and the fermion action is improved by the clover term, the results are quite consistent with the observations for MWT and SU(2) gauge theory with one adjoint Dirac flavor, where a sharp transition appearing at around $\beta=2.0$ has been observed \cite{Catterall:2008qk,Athenodorou:2014eua}. It is important to note that the restrictions appear to be weaker as soon as the fundamental fermions become dynamical. 

\section{Relation to pure $\Nf=2$ SU(2) fundamental}
The SU(2) gauge theory with two fermions in the fundamental representation has been recently investigated in a series of publications, see \cite{Lewis:2011zb,Hietanen:2014xca,Arthur:2016dir} and references therein. As a first important step we must cross-check our simulations by a comparison of our results with these studies. In these references a different lattice action without clover improvement has been chosen. Therefore a comparison in terms of physical units is required. In our studies we measure in addition to the scale $w_0$, the pseudoscalar ($m_{PS}$) and vector ($m_V$) meson masses. The corresponding operators are $\bar{\psi}_1 \gamma_5\psi_2$ and $\bar{\psi}_1 \gamma_k\psi_2$ respectively, where $k$ denotes three different spacial directions. The same operators will be considered later also for the adjoint representation. In that case the additional fermion field is considered in a partially quenched setup described in \cite{Munster:2014cja}. Note that the pseudo-scalar meson is sometimes called pion ($m_\pi$) due to its similarities with the QCD pion state.

\subsection{Scale setting and results in physical units}
In the present studies, we use the parameter $w_0$ obtained from the gradient flow to fix the scale and for the conversion to physical units. The value of $\tau=w_0/a$ is defined from dependence of the action density $E$ on the flow time $\tau$ as the point where the condition
\begin{align}
\tau \frac{d}{d\tau} \tau^2 E(\tau)=W_{ref}
\end{align}
is fulfilled. The flow is taken from the Wilson gauge action and the clover antisymmetric definition is used for $E$ on the lattice. We have chosen a reference scale of $W_{ref}=0.3$ in most cases, which is the common value for scale setting in QCD. As shown in \cite{Bergner:2014ska}, large values of the reference scale are strongly influenced by the large autocorrelation times of topological quantities. Large-$N_c$ analysis supports a scaling with $N_c$, which would lead to $W_{ref}=0.2$ for SU(2) instead of $W_{ref}=0.3$ for SU(3) \cite{Ayyar:2017qdf}. Nevertheless, a reference scale of $W_{ref}=1.0$ has been chosen in \cite{Arthur:2016dir} and we have to consider this value for comparison in physical units. To mark the difference, we denote $w_0/a$ the value at $W_{ref}=0.3$ and $\wt_0/a$ at $W_{ref}=1.0$ in the following. The scales are linearly extrapolated as a function of $(am_{PS})^2$ to the chiral limit ($m_{PS}=0$) to obtain $\wt_{0\chi}/a$ and $w_{0\chi}/a$. At $\beta=2.1$ a linear fit in the range of $(am_{PS})^2<0.3$ yields a value of $\wt_{0\chi}/a=4.17(9)$.
Our limited data for $\beta=2.2$ provide an estimate of $\wt_{0\chi}/a=5.72(12)$.

\subsection{The vector meson of clover improved pure fundamental runs}
\begin{figure}
	\centerline{\includegraphics[width=0.6\textwidth]{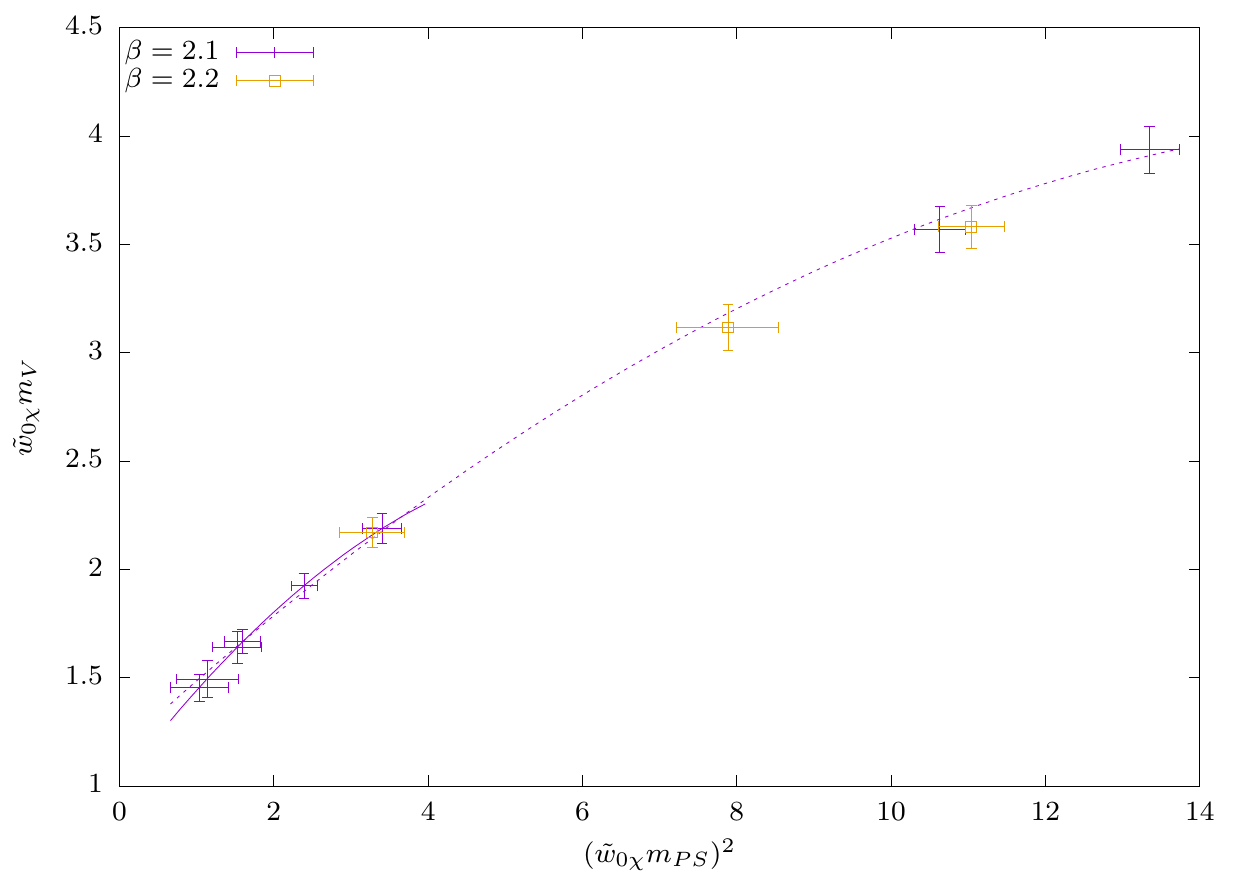}}
	\caption{The vector meson mass of the SU(2) $\Nf=2$ fundamental theory extrapolated to the chiral limit. A quadratic fit function is used for the $\beta=2.1$ data. The dotted line indicates the fit of the complete range, but the final value is obtained only in the range $(w_{0\chi} m_{PS})^2<4$ indicated by the solid line. The $\beta=2.2$ data is added for comparison.  \label{fig:vec} }
\end{figure}
With the common scale setting, the physical value of the vector meson mass in the chiral limit can be compared to previous results for the SU(2) $\Nf=2$ pure fundamental theory. The extrapolation to the chiral limit is done including quadratic corrections as shown in Fig.~\ref{fig:vec}. The fit of the complete range yields a value of $\wt_{0\chi} m_{V\chi}=1.17(4)$. If we restrict the fit range, as done in \cite{Arthur:2016dir}, to $(\wt_{0\chi} m_{PS})^2<4$ the final result is $\wt_{0\chi} m_{V\chi}=1.008(9)$, which is consistent with the continuum extrapolation of \cite{Arthur:2016dir}, $\wt_{0\chi} m_{V\chi}=1.01(3)$. Note that comparing the same $\beta$ value, the deviations from the continuum limit are smaller indicating an improvement by the clover term. The data at $\beta=2.2$ is currently not sufficient to provide a fit in the relevant range, but it is compatible with the fit at $\beta=2.1$. This also indicates that the results are close to the continuum.

\section{Relation to the SU(2) adjoint limit}
\begin{figure}
	\centerline{\includegraphics[width=0.6\textwidth]{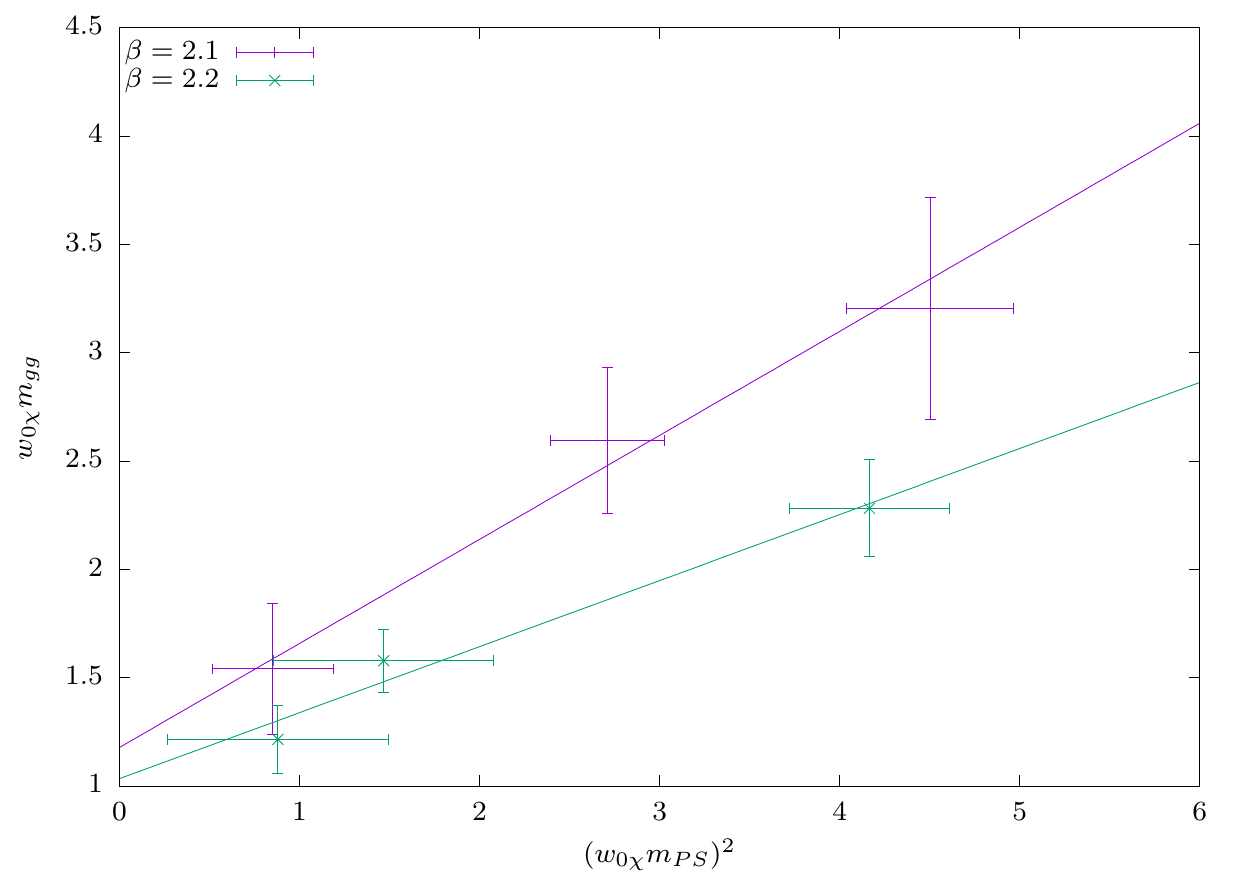}}
	\caption{The mass of the gluino-glue particle in the pure adjoint limit (supersymmetric Yang-Mills theory). The plot shows a linear chiral extrapolation in units of $w_{0\chi}$ as a function of the square of the adjoint pion mass $m_{PS}$.   \label{fig:gg} }
\end{figure}
The SU(2) pure adjoint limit corresponds to supersymmetric Yang-Mills theory (SYM). The simulations with one Majorana fermion require in this limit the RHMC algorithm and have a significantly higher computational cost than the pure fundamental limit. The validation of the results in this limit against our previous data is required since the simulation setup and the lattice action has been changed.

There are no simple physical mesonic states without disconnected contributions in this theory. The adjoint pion mass, corresponding to $m_{PS}$, determines the deviations from the chiral limit using partially quenched chiral perturbation theory \cite{Munster:2014cja}. All particles require rather involved measurements with large statistics, which is inaccessible in this study. The only estimate of a bound state that we can access is the mass of the fermionic gluino-glue particle, see \cite{Bergner:2015adz} for a definition of the operator. This can be compare to our previous results in units of $w_{0\chi}$.

We have generated only a small number of runs and consequently the fit range is insufficient for a precise chiral extrapolation. The rough estimates based on a linear fit are $w_{0\chi}/a=2.42(22)$ at $\beta=2.1$ and around $w_{0\chi}/a=3.85(9)$ at $\beta=2.2$. More precise data would require to consider a larger range of masses since higher order corrections might be relevant. 
The gluino-glue mass in units of the scale $w_{0\chi}$ can be extrapolated to the chiral limit, as shown in Fig.~\ref{fig:gg}. We obtain a value of $w_{0\chi}m_{gg}=1.18(18)$ at $\beta=2.1$ and $w_{0\chi}m_{gg}=1.03(14)$ which is close enough to our previous continuum extrapolation $w_{0\chi}m_{gg}=0.93(6)$ in \cite{Bergner:2015adz}. 

\section{Scenarios for the mixed fundamental-adjoint theory}
In the previous sections, we have confirmed the reliability of our simulations in the limiting pure fundamental and pure adjoint cases. The current first study of the mixed representation theory is organized in such a way that both limiting cases can be reached with the same bare coupling. This means that the range of $\beta$ is limited by the pure adjoint bulk transition. In later studies, even smaller $\beta$ might be considered since the transition is weakened by the fundamental fermions. 

A possible scenario for the theory is a spontaneous chiral symmetry breaking in the limit where both masses tend to zero (chiral limit).
 Like in QCD, Goldstone bosons are expected to interact accordingly to chiral perturbation theory in the small mass regime. Alternatively this theory could be close to an infrared conformal fixed point (IRFP). Near an IRFP, the masses of all particles scale to zero with an exponent provided by the mass anomalous dimension. 
 In a walking or near conformal regime, the behavior could be quite similar, even though it is difficult to quantify the distance to the conformal case. Hence it could be that the theory shows already signs of conformality even if a chiral symmetry breaking scenario is obtained in the deep infrared. 

In the following, we investigate to what extends these two scenarios are reflected in the numerical data. The fact that two different fermion representations are considered leads to some complications, for instance related to the parameter tuning and scale setting, that will be explained in the following. 

\subsection{Parameter-tuning and scale setting}
\begin{figure}
	\subfigure[Dependence on adjoint bare mass\label{fig:mpifm0a}]{\includegraphics[width=0.47\textwidth]{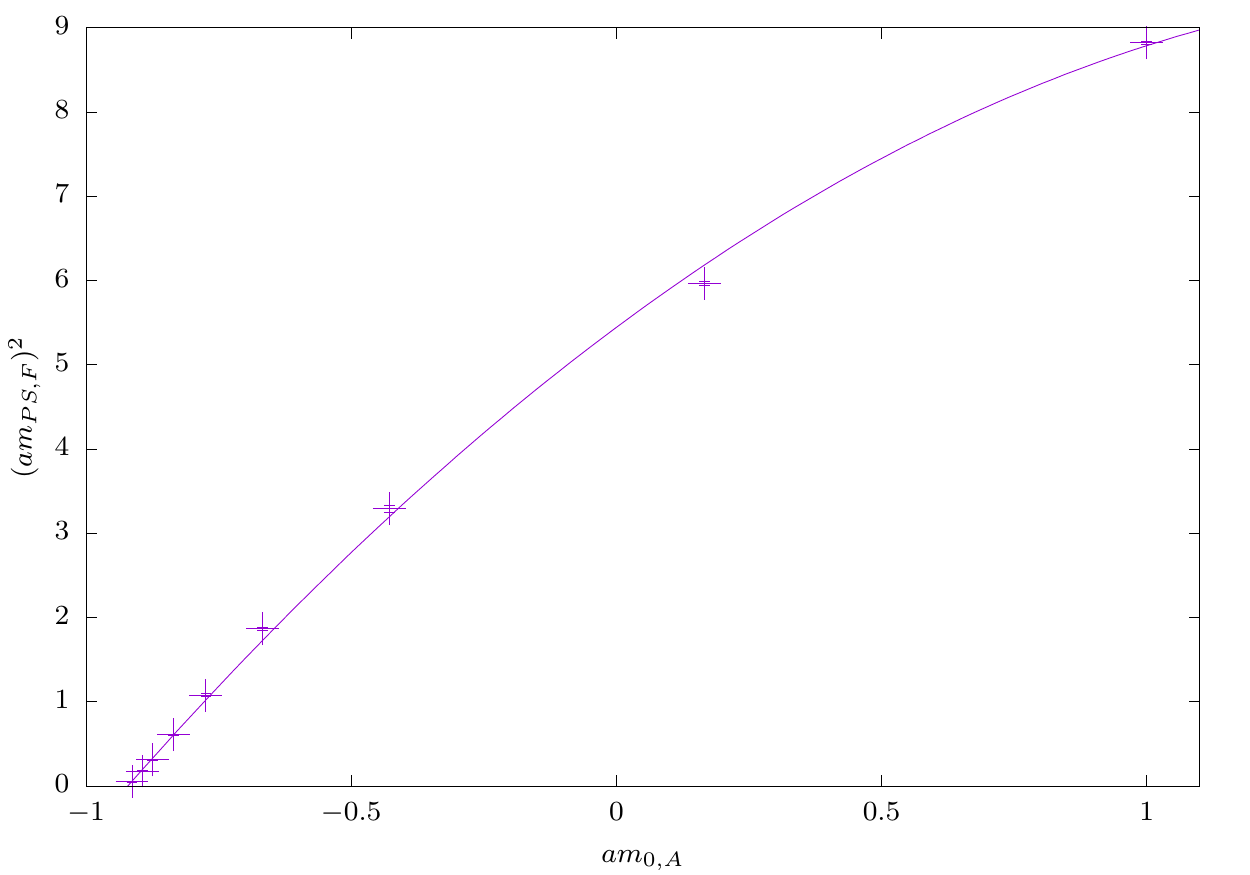}}
	\subfigure[Dependence on fundamental bare mass\label{fig:mpiam0f}]{\includegraphics[width=0.47\textwidth]{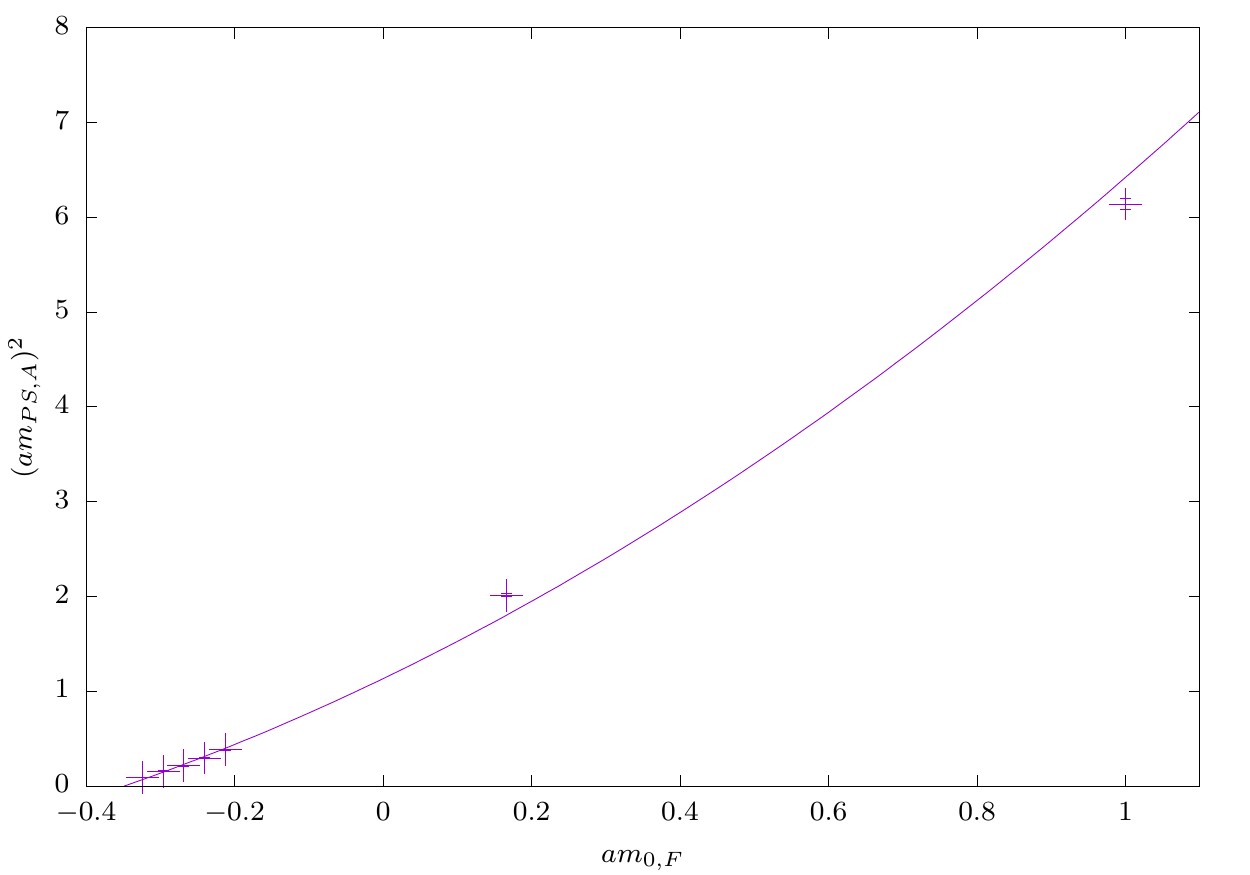}}
	\caption{Dependence of the pseudoscalar mass in one representation on the bare mass parameter of the other. In Fig.~\ref{fig:mpifm0a} $\kf$ is fixed to $0.1350$ in Fig.~\ref{fig:mpiam0f} $\ka$ is fixed to $0.1600$. The quadratic fit is added to illustrate the qualitative trend.\label{fig:m0tun} }
\end{figure}
\begin{figure}
    \centerline{\includegraphics[width=0.7\textwidth]{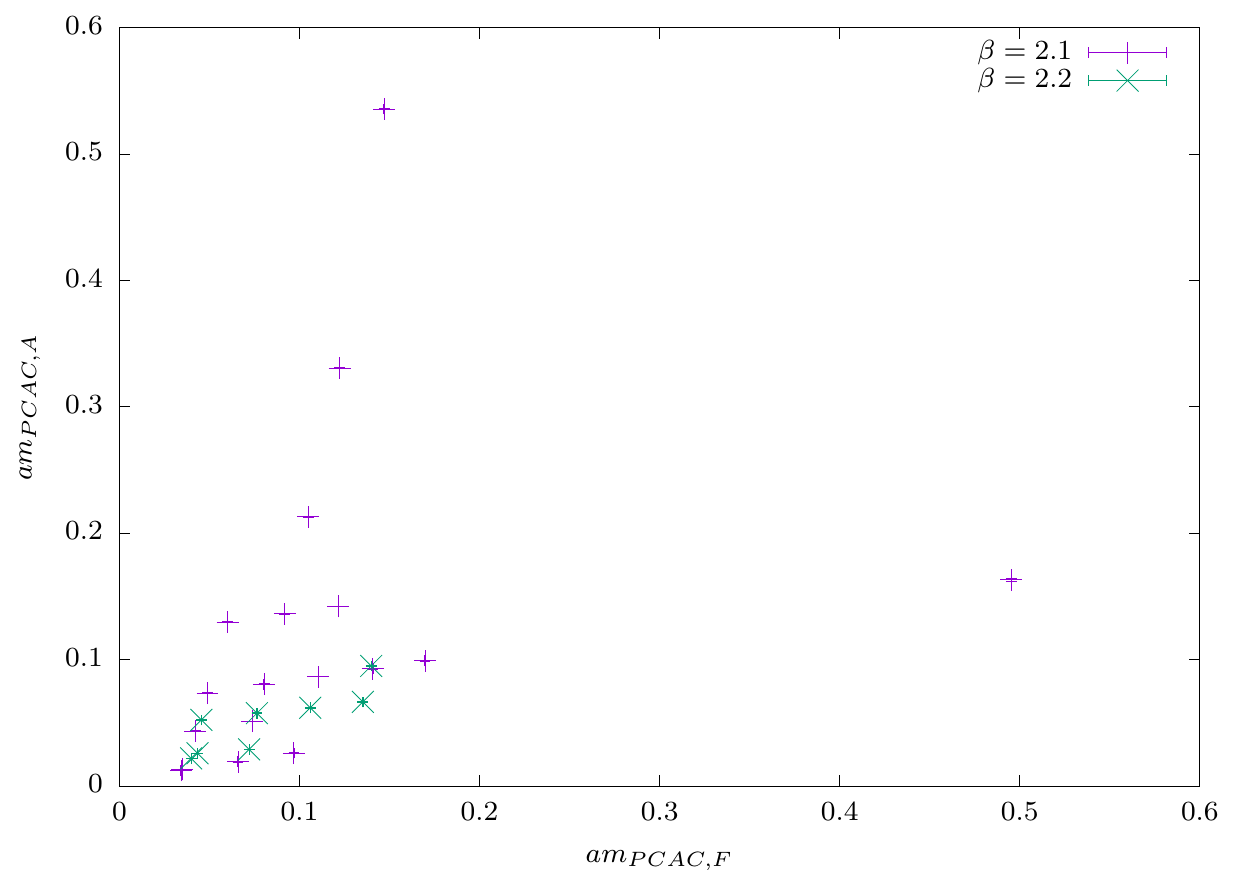}}
	\caption{Range of masses obtained from the partially conserved axial current (PCAC) relation for both of the representations. The adjoint PCAC mass $m_{PCAC,A}$ and the one of the fundamental representation $m_{PCAC,F}$ are shown in lattice units for the two different values of the coupling $\beta$. \label{fig:mrange} }
\end{figure}

The tuning of the bare mass in one representation shows a clear dependence on the mass parameter of the other, as illustrated in Fig.~\ref{fig:m0tun}. Such a dependence complicates the tuning of the theory toward the chiral limit. Note, however, that the bare parameter dependence is not the relevant quantity for fits of chiral perturbation theory in case of Wilson fermions. The mass for the chiral fits is obtained from the partially conserved axial current (PCAC) relation ($m_{PCAC}$). The PCAC masses in both representations for the relevant simulations are shown in Fig.~\ref{fig:mrange}.

Another observation is a significant increase of the flow scale. At $\beta=2.1$, $\kf=0.1360$, $\ka=0.1620$ the value of $w_{0}/a=4.639(77)$ is obtained, which is significantly larger than values at a similar $m_{PS,A}$ in the pure adjoint case (between $w_{0}/a=2.289(11)$ and $3.348(22)$ at the same $\beta$). This indicates a slow running of the coupling. In addition, the topological fluctuations are suppressed, which in turn leads to a large autocorrelation of $w_{0}/a$. Due to this reason and the small statistic, the values for this quantity at $\beta=2.2$ are currently not reliable. A very rough estimate of can be obtained in a chiral extrapolation as a function of the adjoint PCAC mass, which is the dominant dependence compared to the PCAC mass of the fundamental representation. This leads to a prediction of $w_0/a=4.9(2)$ at $\beta=2.1$, see Fig.~\ref{fig:w0ma}. The data at $\beta=2.2$ are within very large errors consistent with this value, but they show a strong correlation with the topological charge. Note that all of our runs are based on a minimum of 300 thermalized configurations. 

\section{Chiral perturbation theory and chiral extrapolations}
\label{sec:chiral}
\begin{figure}
	\subfigure[fundamental representation \label{fig:mpifmf}]{\includegraphics[width=0.47\textwidth]{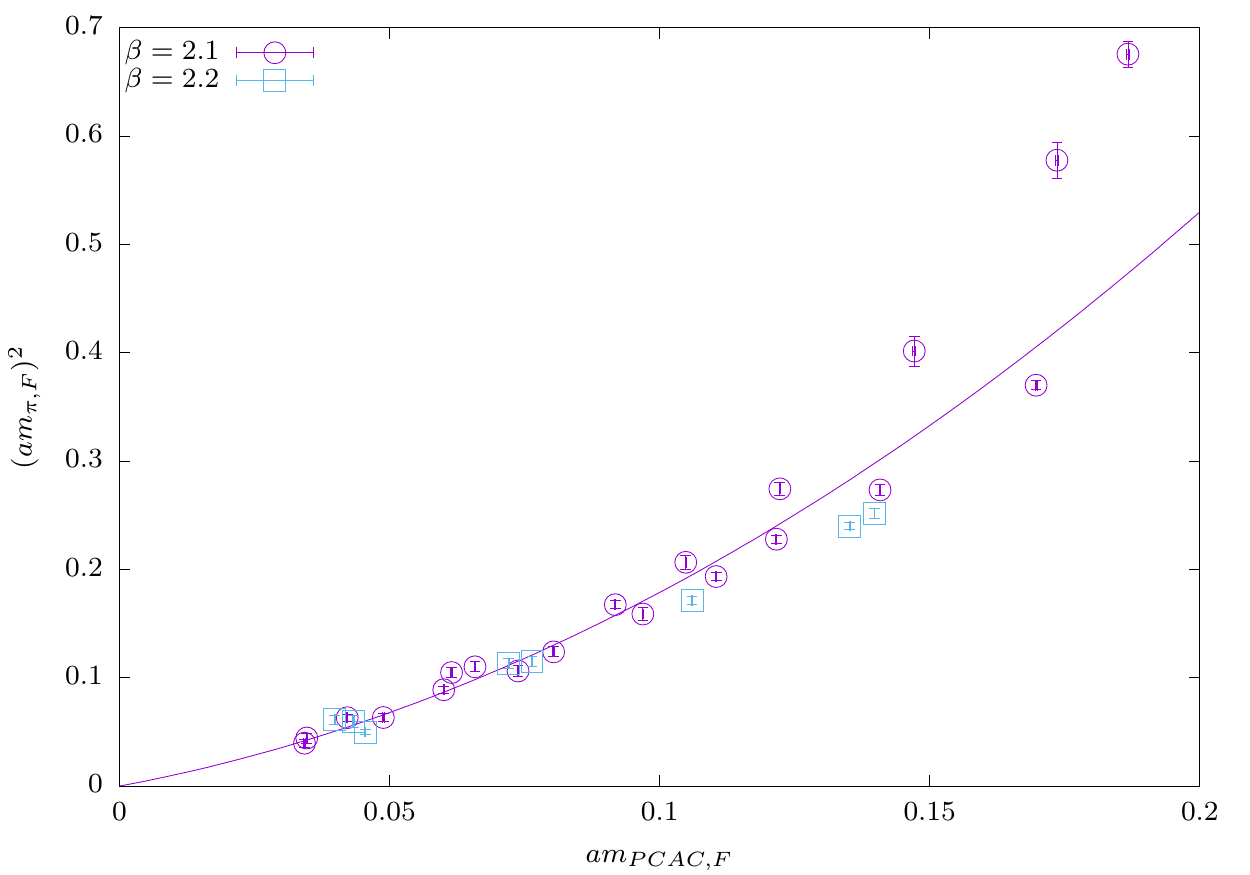}}
	\subfigure[adjoint representation\label{fig:mpiama}]{\includegraphics[width=0.47\textwidth]{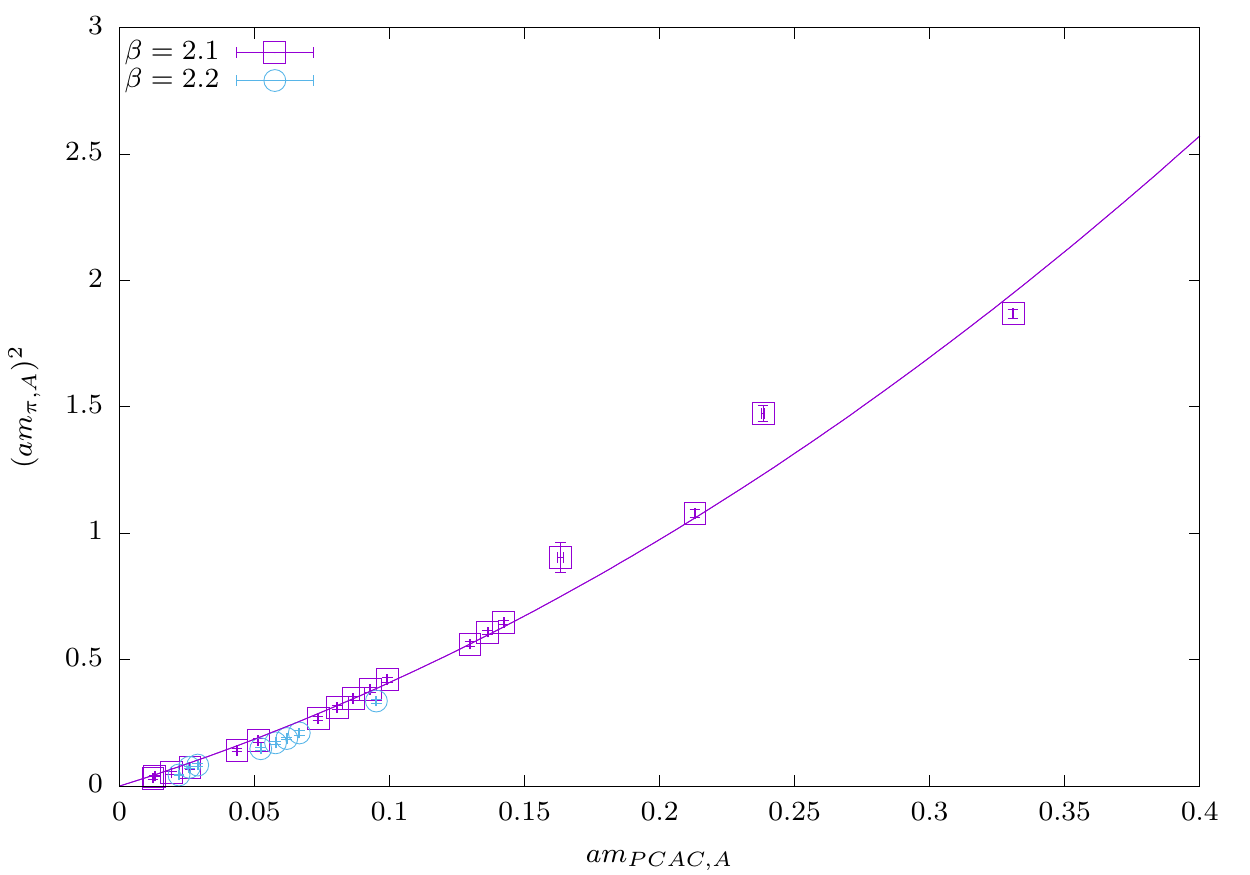}}
	\caption{Dependence of the pseudoscalar mass on the PCAC mass of the same representation. The leading order of chiral perturbation theory predicts a linear behaviour of $m_{PS}^2$. The fit includes an additional quadratic correction, but not the dependence on the mass of the other representation.\label{fig:mpifit} }
\end{figure}
Chiral perturbation theory with two different representations has been worked out in \cite{DeGrand:2016pgq} and applied in the analysis of lattice data in \cite{Ayyar:2017qdf}. As a first step, it is already instructive to test the approximation that neglects the interdependence of the two fermion species. This means to apply a fit function according to $\Nf=2$ chiral perturbation theory for the fundamental representation. The breaking pattern for the gauge group \su{2} is \cite{Lewis:2011zb}
\begin{align}
\su{2\Nff}\rightarrow \text{Sp}(2\Nff)\, ,
\end{align}
leading to three pseudoscalar and two scalar Goldstone states.
The breaking pattern for the adjoint representation is
\begin{align}
\su{2\Nfa}\rightarrow \text{SO}(2\Nfa)\; .
\end{align}
For one adjoint Majorana fermion an approach based on partially quenched chiral perturbation theory has to be used, in which the disconnected part of the pseudoscalar meson (adjoint pion) becomes massless in the chiral limit \cite{Munster:2014cja}. A PCAC mass is measured from the disconnected correlators, which is the same measurement as for $\Nfa=1$, but with gauge configurations of the $\Nfa=1/2$ theory.

We consider first a fit of $(m_{PS})^2$, which is linear in $m_{PCAC}$ at leading order, including also quadratic corrections, see Fig.~\ref{fig:mpifit}. At small $m_{PCAC}$ the fit captures the main dependence, but for the fundamental representation there are considerable differences at larger $m_{PCAC,F}$. This already indicates corrections at the order of the product of the two masses. These corrections seem to be much smaller for the adjoint representation.

In order to include the corrections from the mutual interactions of the fields in the other representation, the fit must include two dimensions. Due to our currently limited data, we consider only the leading contributions in \cite{Ayyar:2017qdf}. Higher order terms are either of logarithmic form or describe lattice artefacts and a determination of the coefficients would require more data and different lattice spacings. The fit functions
\begin{align}
(am_{PS,F})^2&=c_1 am_{PCAC,F}+c_2 am_{PCAC,F}am_{PCAC,A}+c_3 (am_{PCAC,F})^2\\
(am_{PS,A})^2&=c_1 am_{PCAC,A}+c_2 am_{PCAC,F}am_{PCAC,A}+c_3 (am_{PCAC,A})^2\; 
\end{align}
require to determine the unknown parameters $c_1$, $c_2$, and $c_3$. In order to handle finite size effects, we have excluded $m_{PS} L<7$ for both representations. We have tested several different cuts for the fit ranges to optimize the goodness of fit ($\chi^2/\text{dof}$). The final fitting intervals include 11 and 10 points. The fitted parameters are summarized in Tab.~\ref{tab:mpifitres}. The results show that the dominant contribution comes from the mass of the same representation with still a significant mixing with the other representation.
\begin{table}
\begin{center}
\begin{tabular}{|c|ccc|cc|c|}
	\hline
	& $c_1$ & $c_2$ & $c_3$&$am_{PCAC,F}$&$am_{PCAC,A}$&$\chi^2/\text{dof}$\\
	\hline
	$(am_{PS,F})^2$ & 0.902(22) & 1.725(82) & 6.40(18)&<0.2&<0.6&8\\
	$(am_{PS,A})^2$ & 2.941(80) & 2.11(51) & 9.47(50)&<0.5&<0.3&0.8\\
	\hline
\end{tabular}
\end{center}
\caption{Summary of the fit results for the pion masses. The last two columns specify the fit ranges and reduced chi-square. Note that the reduced chi-square without the mass of the second representation ($c_2$) was a factor of 7 (for $m_{PS,F}$) and 4 (for $m_{PS,A}$) larger. \label{tab:mpifitres}}
\end{table}

\begin{figure}
	\subfigure[fundamental representation \label{fig:fpifmf}]{\includegraphics[width=0.47\textwidth]{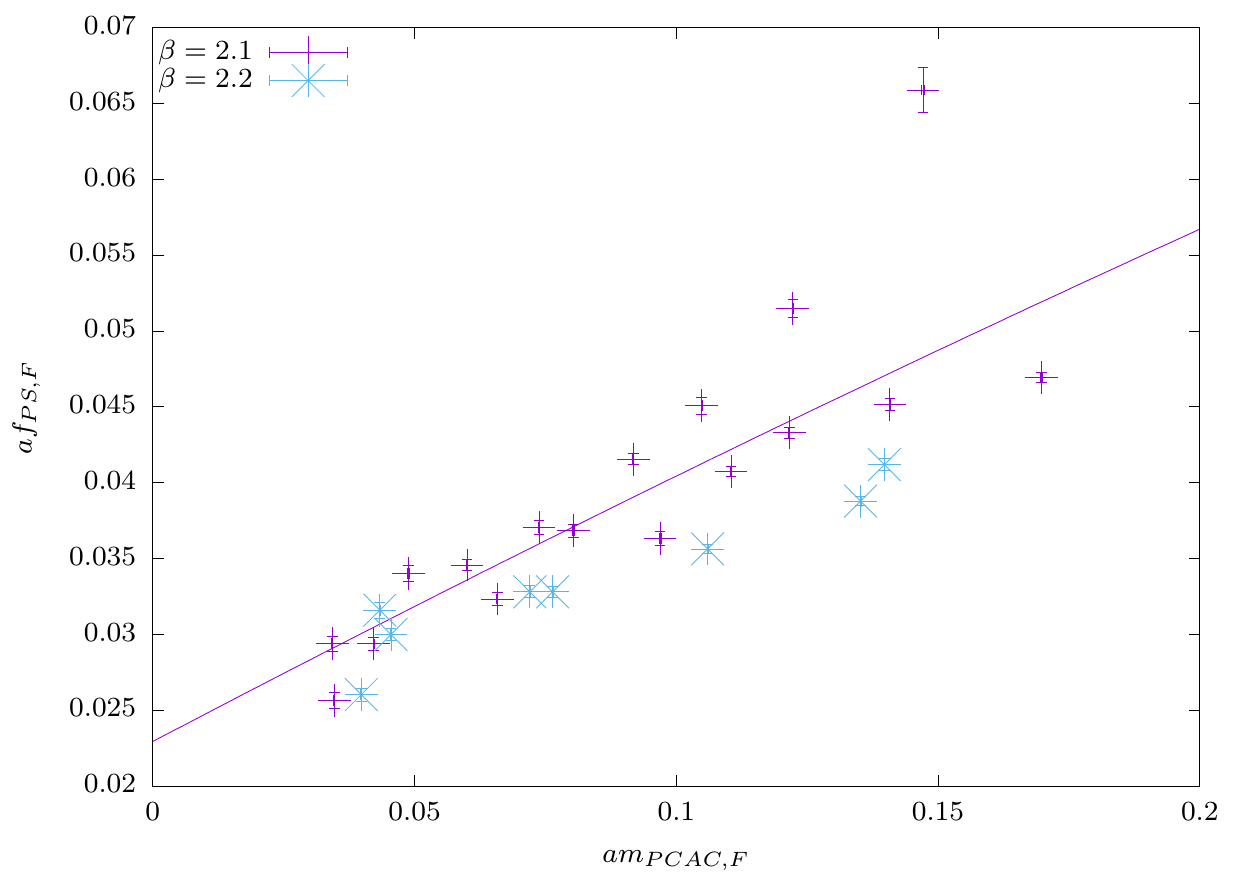}}
	\subfigure[adjoint representation\label{fig:fpiama}]{\includegraphics[width=0.47\textwidth]{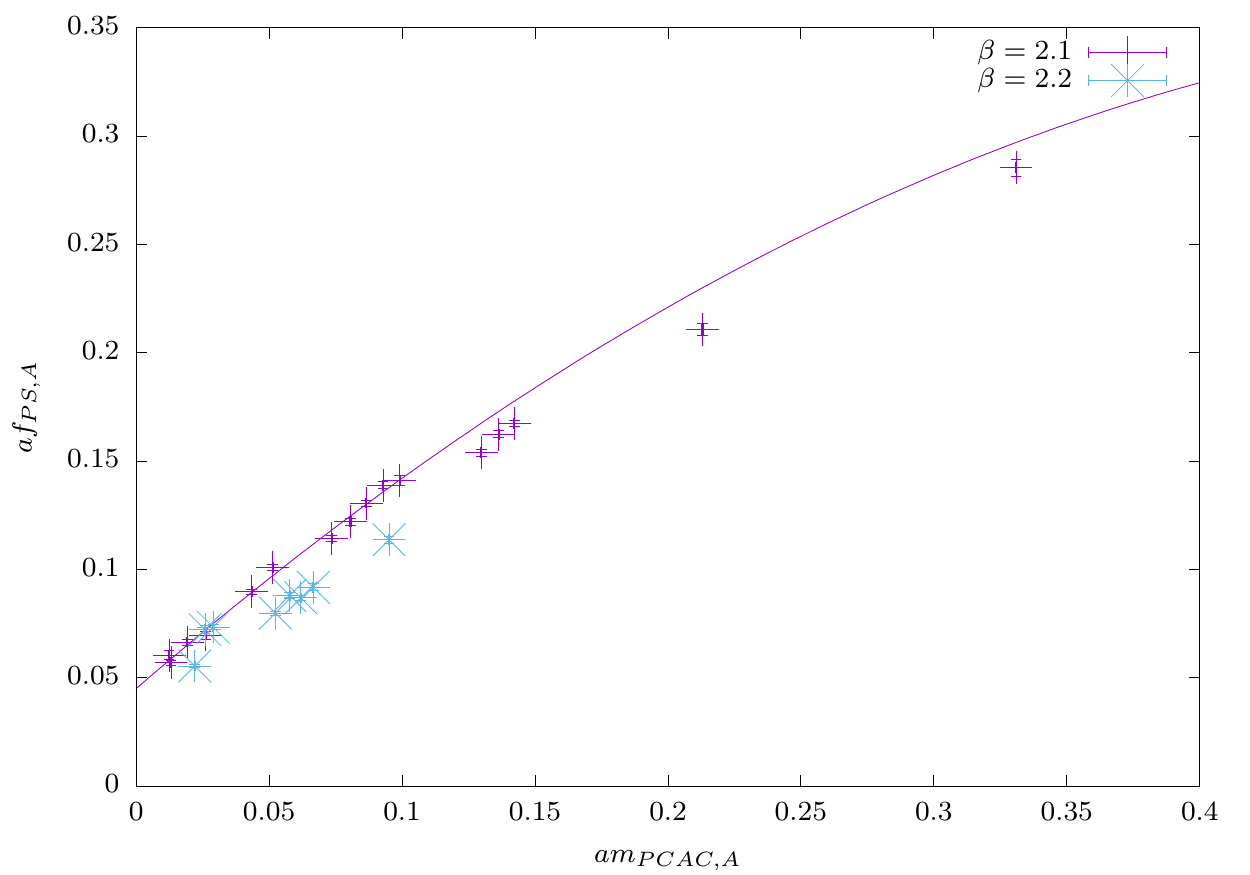}}
	\caption{Dependence of the pseudoscalar decay constant on the PCAC mass of the same representation. Compared to the functions \eqref{eq:fpsf} and \eqref{eq:fpsa}, the fit includes an additional quadratic correction, but not the dependence on the mass of the other representation. The fit in Fig.~\ref{fig:fpiama} is considered in the range $am_{PCAC,A}<0.11$.\label{fig:fpifit} }
\end{figure}
Apart from $m_{PS}$, the chiral symmetry breaking scenario can also be checked with the pseudoscalar decay constant $f_{PS}$. The dependence on the fermion mass in the same representation is shown in Fig.~\ref{fig:fpifit}. In case of the fundamental representation, it indicates clearly an additional dependence on the adjoint mass.
We are again only able to extract the leading behavior of the dependence \cite{Ayyar:2017qdf}
\begin{align}
af_{PS,F}&=c_0+c_1 am_{PCAC,F}+c_2 am_{PCAC,A}\label{eq:fpsf}\\
af_{PS,A}&=c_0+c_1 am_{PCAC,A}+c_2 am_{PCAC,F}\label{eq:fpsa}\;.
\end{align}
This fit resolves the dependence on the two masses, as can be seen in Tab.~\ref{tab:fpifitres}. Note that there is still a rather large residual for the fundamental representation, which could be due to higher order corrections but also to the current limited precision of the data.
\begin{table}
	\begin{center}
		\begin{tabular}{|c|ccc|c|}
			\hline
			& $c_1$ & $c_2$ & $c_3$&$\chi^2/\text{dof}$\\
			\hline
			$af_{PS,F}$ & 0.02465(46) & 0.1139(37) & 0.0387(19)&11\\
			$af_{PS,A}$ & 0.0591(22) & 0.668(15) & 0.116(16)&1.4\\
			\hline
		\end{tabular}
	\end{center}
	\caption{Summary of the fit results for the pseudoscalar decay constants. The last fit ranges are the same as in Tab.~\ref{tab:mpifitres}. Note that the reduced chi-square without the mass of the second representation ($c_2$) is a factor of 5 (for $f_{PS,F}$) and 6 (for $f_{PS,A}$) larger. \label{tab:fpifitres}}
\end{table}

\begin{figure}
	\subfigure[fundamental representation \label{fig:mvfmf}]{\includegraphics[width=0.47\textwidth]{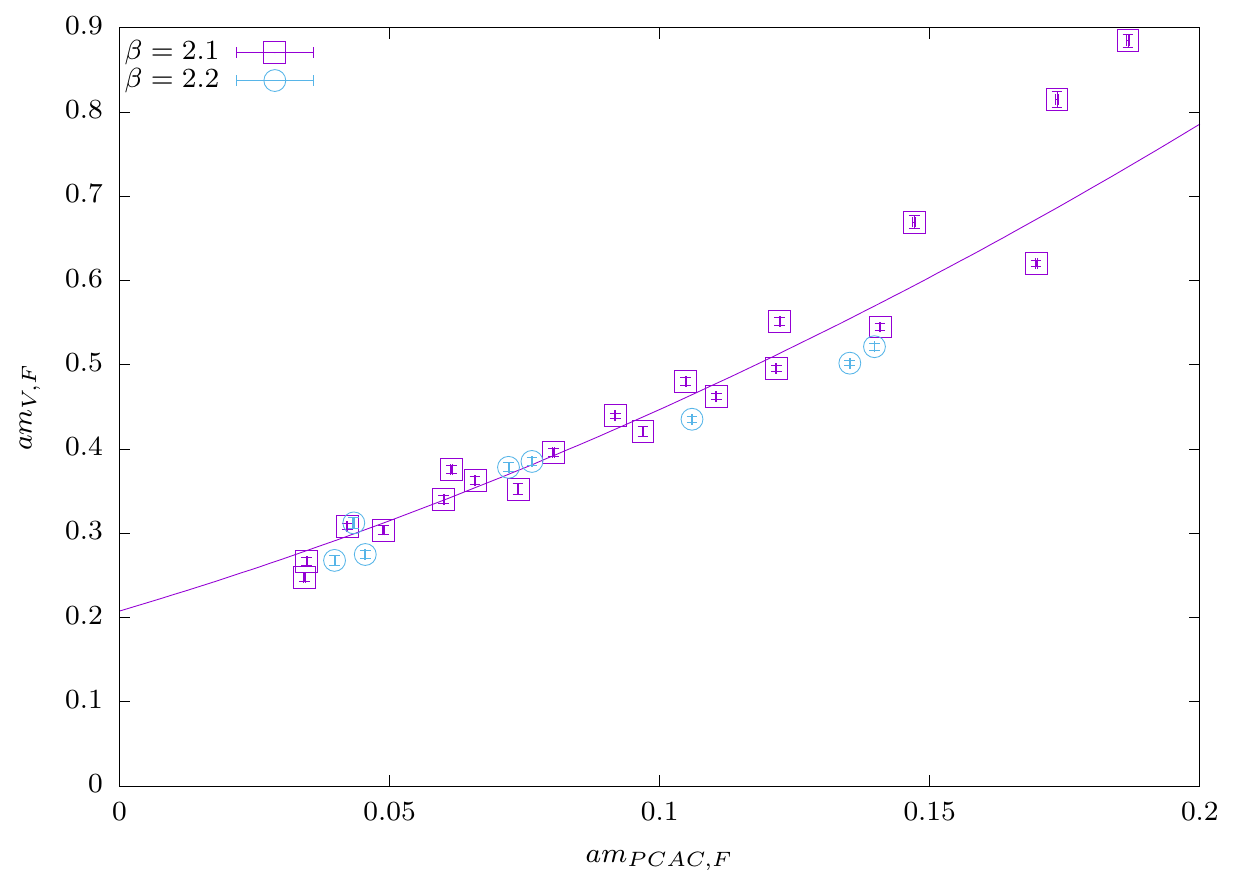}}
	\subfigure[adjoint representaiton\label{fig:mvama}]{\includegraphics[width=0.47\textwidth]{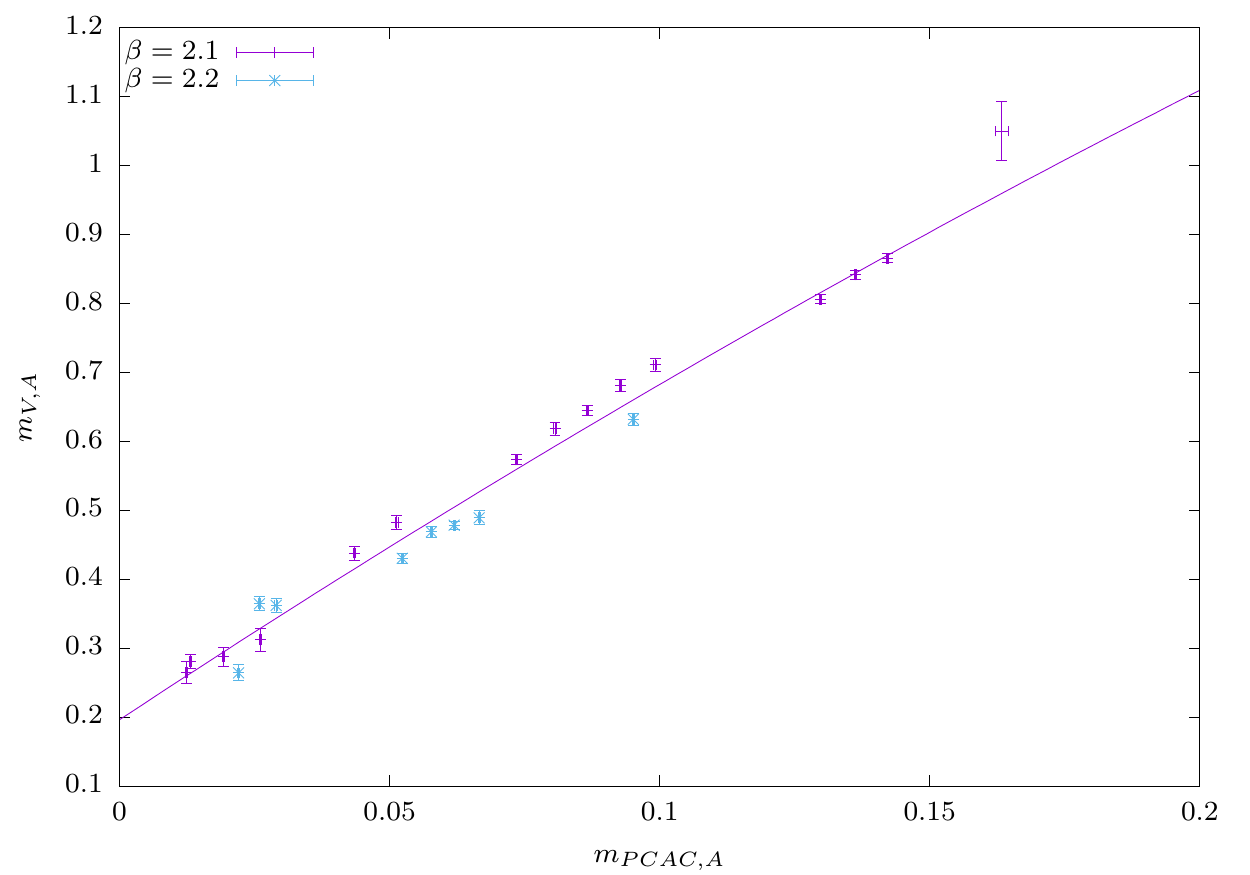}}
	\caption{Dependence of the vector meson mass on the PCAC mass of the same representation. The fit is done with a constant and up to quadratic corrections ignoring the dependence on the mass of the other representation. \label{fig:mvfit}}
\end{figure}

Besides the expected massless states in the scenario of chiral perturbation theory, we can also look at the vector states $m_V$. In case of the adjoint representation, this is again not a physical particle of the theory, but it is obtained in the partially quenched setup considering only the disconnected part of the correlator\footnote{Note that the vector meson is not even a physical state in adjoint QCD with one Dirac fermion, but it is a physical state for $\Nfa=2$.}. It is extrapolated to a constant value in the chiral limit and we consider linear and quadratic corrections in the fit.  As shown in Fig.~\ref{fig:mvfit}, there are again considerable deviations in case of the fundamental representation, whereas for the adjoint representation the dependence in $m_{PCAC,A}$ seems to be dominant. The mass of the gluino-glue particle $m_{gg}$ can also be extrapolated to the chiral limit as a function of $m_{PCAC,A}$, but due to the limited statistics there are much larger uncertainties and the errors are currently underestimated.

\begin{figure}
	\subfigure[Gluino-glue mass\label{fig:mggma}]{\includegraphics[width=0.47\textwidth]{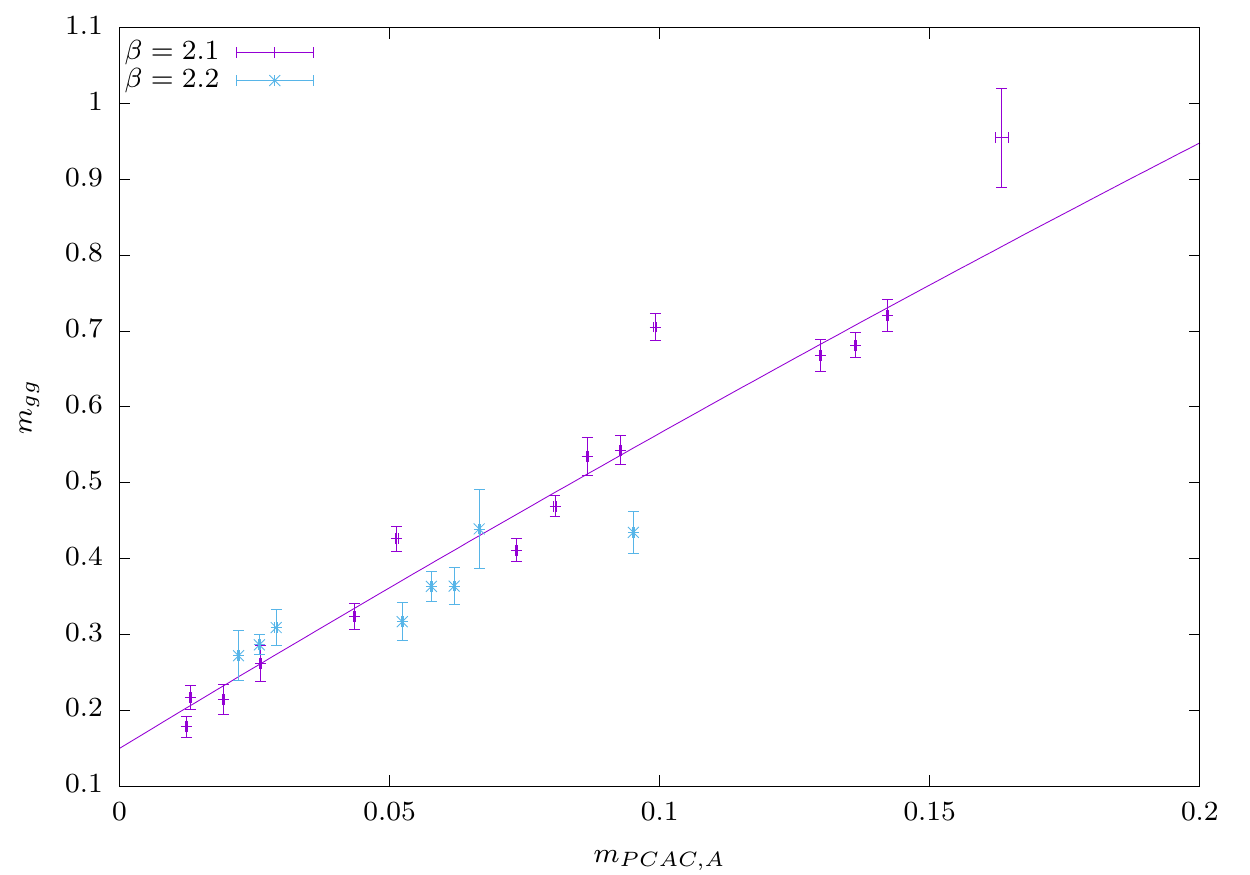}}
	\subfigure[$w_0/a$ chiral extrapoltation \label{fig:w0ma}]{\includegraphics[width=0.47\textwidth]{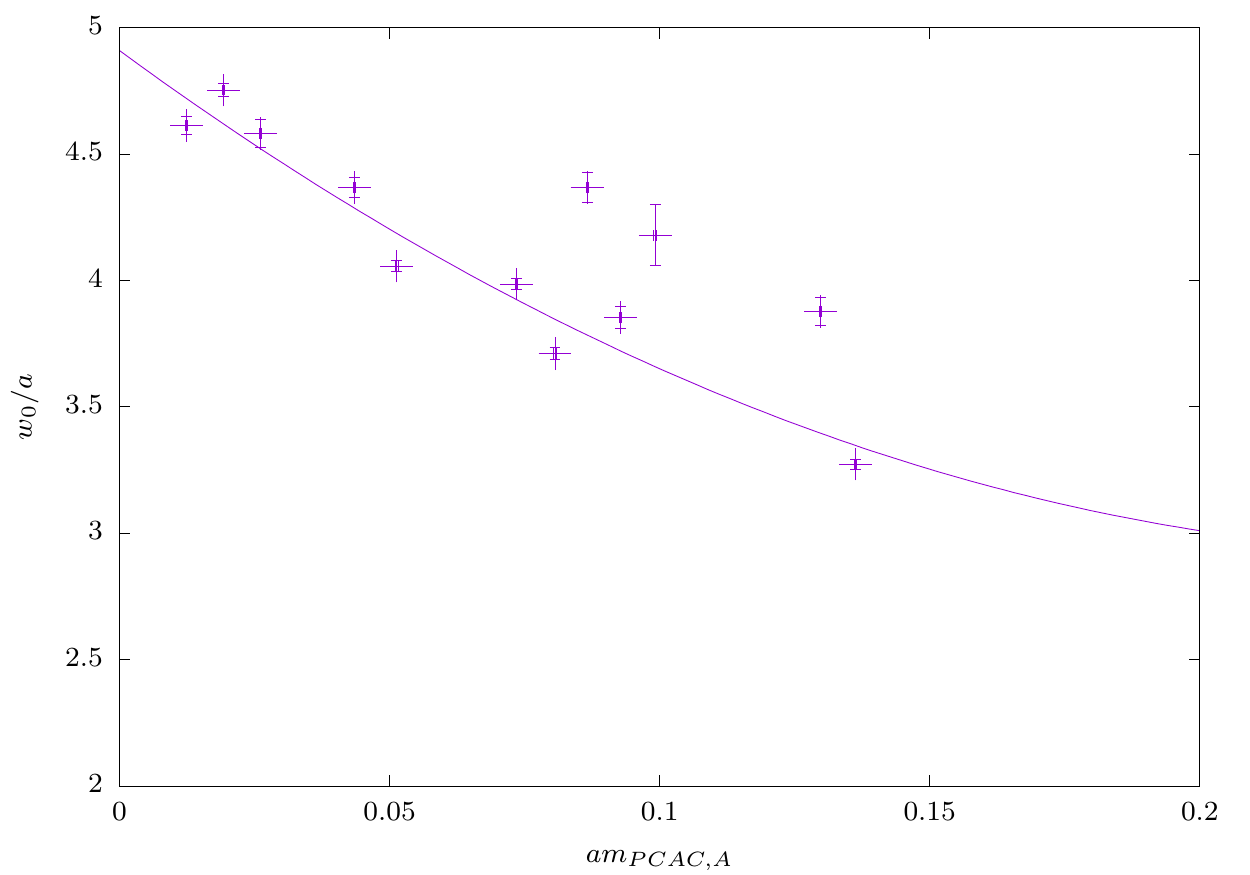}}
	\caption{Chiral extrapolations of the gluino-glue mass and the scale $w_0/a$ as a function of the adjoint PCAC mass, neglecting the dependence on the fundamental mass. Only the data for $\beta=2.1$ is shown in Fig.~\ref{fig:w0ma}. }
\end{figure}
A generic fit ansatz for $m_V$ (and $m_{gg}$) are the following functions
\begin{align}
am_{V,F}&=c_0+c_1 am_{PCAC,F}+c_2 am_{PCAC,A}\\
am_{V,A}&=c_0+c_1 am_{PCAC,A}+c_2 am_{PCAC,F}\\
am_{gg}&=c_0+c_1 am_{PCAC,A}+c_2 am_{PCAC,F}\, .
\end{align}
We have tested higher order corrections, but it turned out that they are not significant and cannot be reliably estimated in this case. We have considered the same fit ranges as for $m_{PS}$. The results are summarized in Tab.~\ref{tab:mvfitres}. The dominant contribution comes still from the fermion mass in the same representation. In case of the adjoint representation, it is not even clear that the inclusion of the mass of the fundamental representation provides a more reliable fit. The gluino-glue mass is clearly dominated by the fermion mass of the adjoint representation and the inclusion of the fundamental representation seems not necessary given the current uncertainties.
\begin{table}
	\begin{center}
		\begin{tabular}{|c|ccc|c|}
			\hline
			& $c_0$ & $c_1$ & $c_2$&$\chi^2/\text{dof}$\\
			\hline
			$am_{V,F}$ & 0.1625(50) & 2.545(40) & 0.241(13)&4\\
			$am_{V,A}$ & 0.288(11) & 3.825(70) & 0.258(67)&2\\
			$am_{gg}$ & 0.213(20) & 2.82(16) & 0.80(13)&6\\
			\hline
		\end{tabular}
	\end{center}
	\caption{Summary of the fit results for the pion masses. The last two columns specify the fit ranges and reduced chi-square. Note that the reduced chi-square without the second representation ($c_2$), but including instead a quadratic correction, is a factor of 10 larger for $m_{V,F}$. For $m_{V,A}$ the reduced chi-square is approximately the same and for $m_{gg}$ it is larger by a factor of 2. \label{tab:mvfitres}}
\end{table}

We have also done a number of simulations at a larger value of $\beta$ equal to 2.2. In the Fig.~\ref{fig:mpifit}, Fig.~\ref{fig:fpifit}, and Fig.~\ref{fig:mggma} we have added these data to the fit at $\beta=2.1$. Compared to the other uncertainties, only a rather small dependence on the gauge coupling can be observed. From the present data, we are not able to resolve a continuum extrapolation.

\section{Conformal scenario}
In a conformal scenario, the behavior of the theory is influenced by an infrared fixed-point (IRFP) of the massless theory. If the scalar matter fields and their interactions were included in the Lagrangian of our theory, the corresponding SQCD theory has $N_c=2$, $\Nf=2$ and is conjectured to be outside the conformal window \cite{Seiberg:1994pq}. Naively the scalar matter-fields are expected to decrease the running of the coupling ($\beta$-function) and bring the evolution toward an IRFP. Hence the conformal scenario is not expected in our theory with $\Nff=2$ and $\Nfa=1/2$. Nevertheless it is interesting to check for a walking (near conformal) scenario, which has been conjectured for the theory with $\Nfa=1$ based on perturbative estimates. In such a scenario it is assumed that in a certain range of scales the $\beta$-function is already strongly influenced by an IRFP appearing at a larger $\Nf$. In this range of scales it should be close to a conformal scenario. We therefore test the scaling of the masses to the chiral point assuming the presence of an IRFP. More generally, the difference between a chirally broken or a conformal scenario is quite subtle for theories close to the lower edge of the conformal window, and we also want to investigate how well the two scenarios can be distinguished.

In the conformal case, the masses are relevant directions of the renormalization group transformation in the vicinity of the IRFP, while the gauge coupling is irrelevant. If one considers $N_f$ fermions in one single representation, each mass is an independent relevant direction. However, they all have a common scaling dimension and it is therefore rather trivial to consider the RG flow in the manifold spanned by the mass parameters. The theory defined by a certain RG trajectory is determined by the ratios of the different mass parameters. In a simplified setup one can consider just two relevant directions defined by the mass of $(\Nf)_h$ heavy and $(\Nf)_l$ light flavours. The flow is in this case defined by the ratio of the two masses. The mass parameters can be used to tune the running of the gauge coupling to a walking scenario. Projected onto the plane of the gauge coupling, the two mass parameters introduce an intermediate region of scales with a small running of the gauge coupling. This is the basic idea of the scenario proposed in \cite{Brower:2015owo,Hasenfratz:2016gut}.

The situation becomes more complicated for a theory with two different representations, since the anomalous dimensions of the two relevant mass parameters are not the same. We therefore recall some of the arguments presented in \cite{DelDebbio:2010ze,Brower:2015owo,Hasenfratz:2016gut} and extend them to the current case. When the scale is changed by $\mu\rightarrow \mu/b$ ($b>1$), the scaling of a correlation function near the fixed point is given by
\begin{align}
C_H(t,g_i,m_i,\mu)=b^{-2y_H}C_H(t/b,b^{y_{g_i}}g_i,b^{y_i}m_i,\mu/b)\,,
\end{align}
with scaling exponents of the operator $y_H$ and the relevant mass directions with scaling dimension $y_i$. The irrelevant couplings $g_i$ with scaling dimension $y_{g_i}<1$ can be approximately set to their fixed point value and are ignored in the following. We can assume that the scale is set by one of the mass parameters $b=m_1^{-1/y_1}$, while the other fermions are below the decoupling scale ($m_1^{-y_i/y_1}m_i$ is small enough).
The scaling is simplified to ($i\neq 1$)
\begin{align}
C_H(t,m_1,m_i)=m^{-2y_H/y_1}C_H(tm_1^{1/y_1},m_i/(m_1^{y_i/y_1}))\; .
\end{align}
For large $t$ the correlation function approaches the exponential form $\exp(-M_Ht)$ with the mass of some state $M_H$. Consequently all particle masses should scale like $m_1^{1/y_1}$ and the dependence on other mass parameters is via the ratio $m_i m_1^{-y_i/y_1}$, or equivalently $m_1 m_i^{-y_1/y_i}$. 
As shown in \cite{Hasenfratz:2016gut}, mass ratios can be considered to cancel the leading $m_1^{1/y_1}$ dependence.

In our case, we can consider the ratio of vector and pseudoscalar mesons, for example, in order to check whether it can be represented by a functional dependence $F_R$ given by 
\begin{align}\label{eq:confscale}
\frac{am_{V,F}}{am_{PS,F}}=F_R(am_{PCAC,F} (am_{PCAC,A})^{-y_F/y_A})\; .
\end{align}
This study requires the determination of anomalous dimensions $\gamma_F$ and $\gamma_A$ for adjoint and fundamental representation, which define the scaling $y_i=1+\gamma_i$.

\begin{figure}
	\subfigure[fundamental representation \label{fig:confloglogf}]{\includegraphics[width=0.47\textwidth]{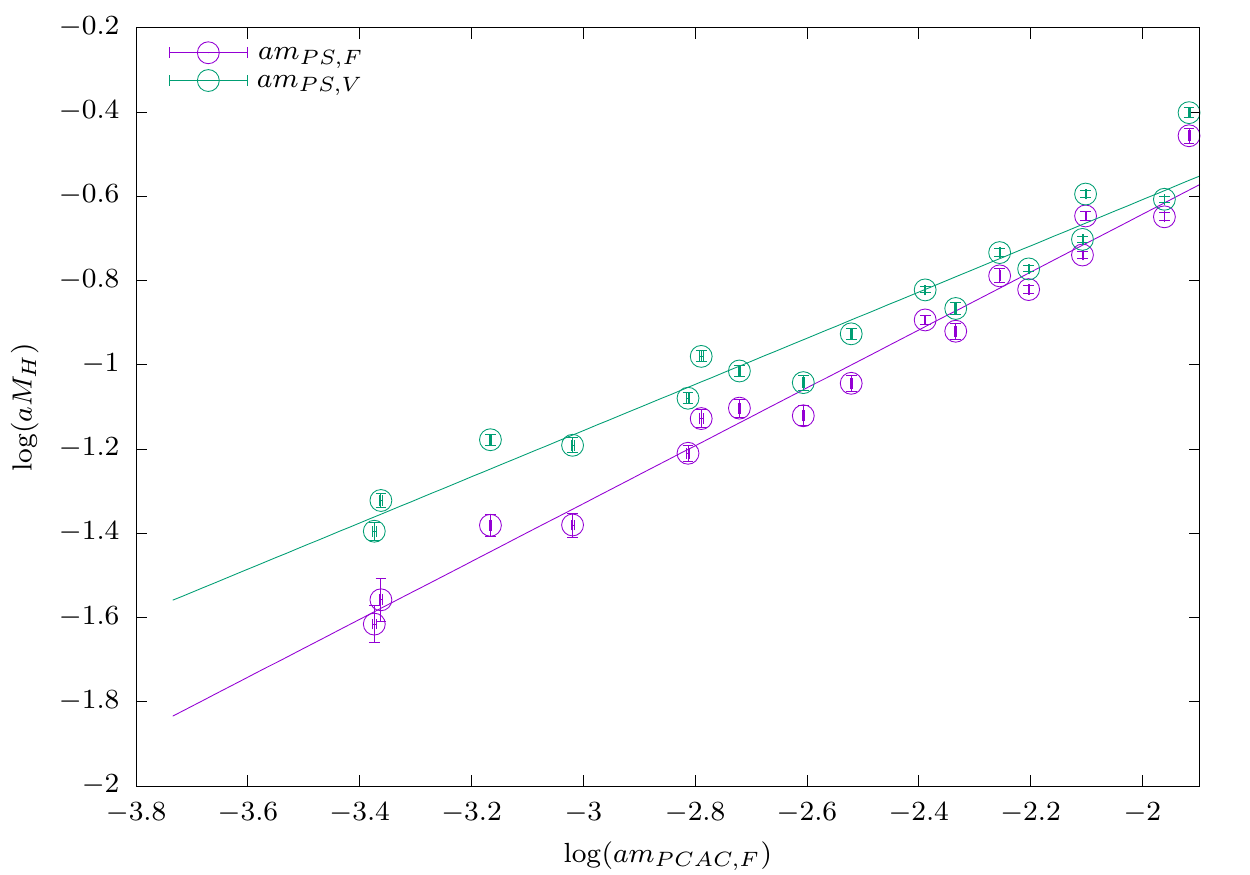}}
	\subfigure[adjoint representaiton\label{fig:conflogloga}]{\includegraphics[width=0.47\textwidth]{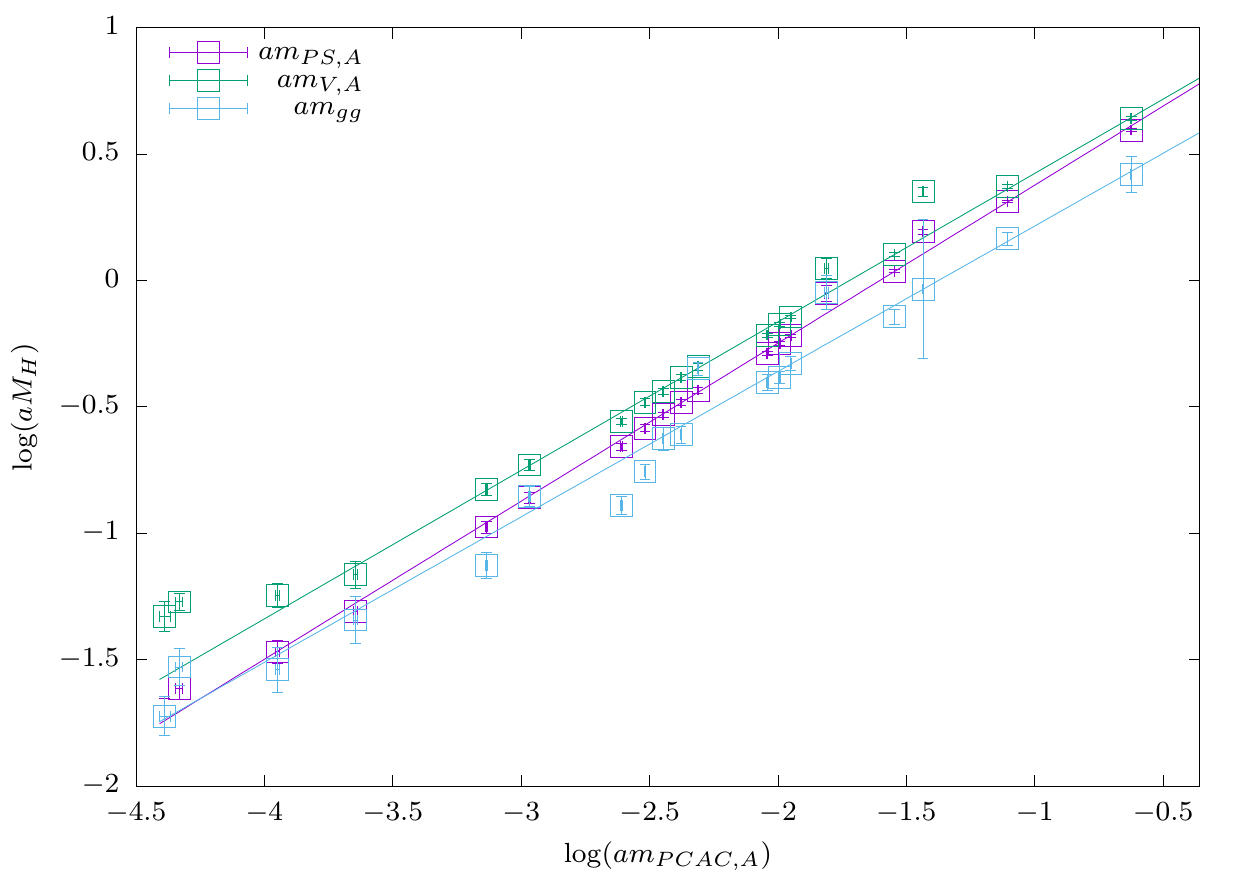}}
	\caption{Double log representation of the dominant dependence of the bound state masses on the two mass parameters. The linear fit assumes an approximate scaling near an IRFP. \label{fig:confloglog}}
\end{figure}
For this reason, let us first investigate the leading dependence $M_H\sim m_i^{1/(1+\gamma_i)}$. Therefore we consider for each state only the leading dependence, either $m_1=m_f$ or $m_2=m_a$, already determined in the chiral fits of Sect.~\ref{sec:chiral}. 
The simplest way to extract the leading exponents is a linear fit in a double logarithmic representation. 
We have done this using $am_{PS,F}$ and $am_{V,F}$ as a function of $am_{PCAC,F}$ (Fig.~\ref{fig:confloglogf}). The $am_{V,F}$ mass shows a rather large deviation from the expected scaling which might be due to the additional dependence on $am_{PCAC,A}$. A fit of the mass $am_{PS,F}$ leads to a scaling dimension $y_F=1.46(8)$. The fit of $am_{PS,A}$, $am_{V,A}$, and $am_{gg}$ shows a more consistent scaling as a function of $am_{PCAC,A}$ (Fig.~\ref{fig:conflogloga}). The obtained scaling dimensions are  $y_A=1.60(2)$ for $am_{PS,A}$, $y_A=1.70(4)$ for $am_{V,A}$, and $y_A=1.26(12)$ $am_{gg}$. We have also used the mode number with the same methods as in \cite{Bergner:2017gzw} to estimate the mass anomalous dimension for the run with the smallest masses for both representations. Since we have not done a careful analysis of the mass dependence, only a rough estimate can be provided from this method, which is $\gamma_F\approx 0.3$ and $\gamma_A\approx 1.0$. In summary these investigations of the leading dependence indicate a $\gamma_F$ in the range $0.3$ to $0.5$ and $\gamma_A$ in the range $0.3$ to $1.0$, assuming conformal scaling.

\begin{figure}
	\centerline{\includegraphics[width=0.47\textwidth]{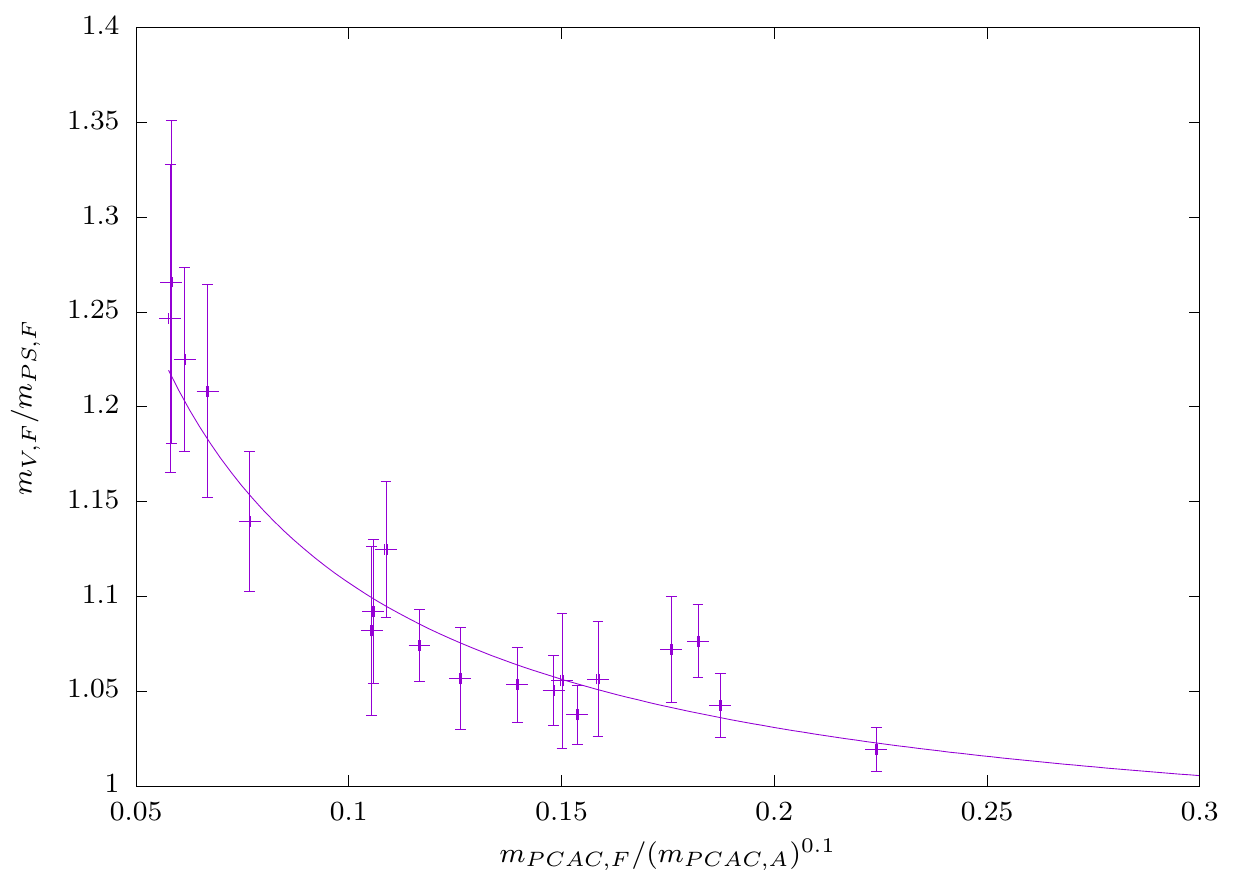}}
	\caption{The ratio of vector over pseudscalar mass in the fundamental representation as a function of $am_{PCAC,F} (am_{PCAC,A})^{-r}$. The line shows a fit according to $a+b/x$. The value of $r=0.1$ corresponds to an optimal choice according to the residual. \label{fig:confrat}}
\end{figure}
The leading scaling can be used to investigate finally the sub-leading dependence \eqref{eq:confscale}. We have investigated the ratio of $m_{V,F}/m_{PS,F}$ as a function of $am_{PCAC,F} (am_{PCAC,A})^{-r}$. The fit according to $a+b/(am_{PCAC,F} (am_{PCAC,A})^{-r})$ can be used to optimize the residual as a function of $r$. In this process, we observe that small values of $r<0.4$ are preferred and a value of $r=0$ cannot be excluded, see Fig.~\ref{fig:confrat}, indicating rather a chiral symmetry breaking than a conformal scenario. Note, however, that the ratio $m_{V,F}/m_{PS,F}$ still shows only a weak dependence on the masses and the consistent scaling in Fig.~\ref{fig:confrat} leads to almost constant ratios between the masses in this figure.

\section{Conclusions}
We have presented the first results of the simulations with a gauge theory coupled to fermions in adjoint and fundamental representation. The present work provides a preparatory study for further simulations of SQCD, UMWT, or compactified SYM with fundamental matter. In the current study we have considered SU(2) gauge theory with one Majorana fermion in the adjoint representation and two fundamental flavours. This corresponds to two flavor SU(2) SQCD without scalar fields.

We have presented cross-checks with existing results in the pure adjoint and fundamental limit. In a first investigations we have also determined the reliable range of parameters by an investigation of the bulk transition. Our investigation shows that the theory is consistent with a chiral symmetry breaking scenario, but still close to the conformal window. The running of the gauge coupling is quite small and mass ratios are nearly constant. This situation is similar to what is expected for SU(2) SQCD with two flavors, since the conformal window is predicted to start close by at $\Nf=\frac{3 N_c}{2}$.  The number of colors $N_c$ would have to be increased in order to get a significative difference from the conformal case with two flavor SU($N_c$) SQCD.

The bare parameter tuning with Wilson fermions of one representation seems to be considerably affected by the other. The dependence on physical mass parameters is provided by the two PCAC masses. We observe a rather mild dependence of the states in one representation on the mass of the other. The most significant effect has the adjoint fermion mass since the fundamental meson masses clearly depend on it. Furthermore, the adjoint mesons and the gluino-glue can be extrapolated to the chiral limit considering as a good approximation only the dependence on the adjoint mass.

Regarding the simulations of SU(2) two flavor SQCD, the following strategy could be derived from the data. Since the system is close to the conformal window, one might neglect the running of marginal couplings like the gauge coupling and set them to their tree level value. The important extrapolation of the pure SYM part to the chiral limit can be done as a first approximation just from the adjoint PCAC mass, which might capture the leading dependence. Note, however, that due to Yukawa couplings the chiral transformations of fundamental and adjoint fermions are not independent in SQCD. The Yukawa terms are only invariant under a combined transformation of the two fermion fields and the scalar field.

In order to investigate in more detail the conformal scenario with two different representations, a larger number of adjoint fields should be considered. One interesting investigation is the simulation of UMWT, which means  one Dirac instead of one Majorana fermion in the adjoint representation.

The current data correspond only to a first investigation. A larger statistic, a more complete scan of the relevant parameter range, and a more careful considerations of methods such as the mode number measurement for two fermion representations would be required to provide more precise data. Note, however, that the parameter scan with two independent fermion representations will require a considerably larger amount of computational resources.

\section*{Acknowledgments}
The authors gratefully acknowledge the Gauss Centre for Supercomputing e.V. (www.gauss-centre.eu) for funding this project by providing computing time on the GCS Supercomputer SuperMUC at Leibniz Supercomputing Centre (www.lrz.de). Further computing time has been provided on the compute cluster PALMA
of the University of M\"unster. G.~Bergner acknowledges support from the Deutsche Forschungsgemeinschaft (DFG) Grant No.~BE 5942/2-1.
\begin{appendix}
\section{Summary of the data: pure adjoint and pure fundamental runs}
	\begin{center}
		\begin{small}
		  \input{tablepurefund}
		  \input{tablepureadj}
		\end{small}
	\end{center}
This table is a summary of the data of the pure adjoint ($\kappa_F=0$) and pure fundamental ($\kappa_A=0$) runs. The clover coefficient was set to the one loop values of $C_{SW,F}=1.297$ and $C_{SW,A}=1.696$ at $\beta=2.1$ ($C_{SW,F}=1.283$ and $C_{SW,A}=1.664$ at $\beta=2.2$).
\section{Summary of the runs with two dynamical representations}
\resizebox{!}{0.95\textheight}{
\begin{adjustbox}{angle=90}
\begin{minipage}{\textheight}
\begin{tiny}
				\input{tablepuremixed}
\end{tiny}\\
This table summarizes the runs with two dynamical representations. The clover coefficient was again set to the one loop values of $C_{SW,F}=1.297$ and $C_{SW,A}=1.696$ at $\beta=2.1$ ($C_{SW,F}=1.283$ and $C_{SW,A}=1.664$ at $\beta=2.2$).
\end{minipage}
\end{adjustbox}}
\end{appendix}

\end{document}

%% file: tablepurefund.tex
\begin{tabular}{llrllll}
\toprule
$\beta$ & $N_s\times N_t$ &  $\kappa_F$ & $am_{PS,F}$ &  $am_{V,F}$ &     $w_1/a$ & $N_{conf}$ \\
\midrule
 2.1 &  $24\times 48$ &  0.1350 &  0.8762(69) &  0.9443(58) &  2.3953(50) &  587 \\
 2.1 &  $24\times 48$ &  0.1360 &  0.7819(63) &  0.8559(71) &  2.5433(64) &  502 \\
 2.1 &  $24\times 48$ &  0.1390 &  0.4423(61) &  0.5249(49) &  3.287(12) &  800 \\
 2.1 &  $24\times 48$ &  0.1395 &  0.3709(42) &  0.4614(43) &  3.426(12) &  2600 \\
 2.1 &  $24\times 48$ &  0.1400 &  0.2964(87) &  0.3937(93) &  3.753(30) &  684 \\
 2.1 &  $24\times 48$ &  0.1403 &  0.256(11) &  0.358(12) &  3.878(26) &  673 \\
 2.1 &  $32\times 48$ &  0.1403 &  0.244(10) &  0.3485(78) &  3.972(23) &  749 \\
 2.2 &  $24\times 48$ &  0.1350 &  0.5808(52) &  0.6265(42) &  3.7732(44) &  860 \\
 2.2 &  $24\times 48$ &  0.1360 &  0.4909(92) &  0.5450(75) &  3.876(48) &  350 \\
 2.2 &  $24\times 48$ &  0.1375 &  0.3164(61) &  0.3796(41) &  4.952(51) &  795 \\
 2.2 &  $32\times 64$ &  0.1375 &  0.3237(55) &  - &  - &  1144 \\
\bottomrule
\end{tabular}

%% file: tablepureadj.tex
\begin{tabular}{llrllll}
\toprule
$\beta$ & $N_s\times N_t$ &  $\kappa_A$ & $am_{PS,A}$ &  $am_{gg}$ &     $w_0/a$ & $N_{conf}$ \\
\midrule
 2.1 &  $24\times 48$ &  0.1680 &  0.877(11) &  1.324(91) &  1.3182(15) &  320 \\
 2.1 &  $24\times 48$ &  0.1690 &  0.6803(97) &  1.073(41) &  1.6071(18) &  302 \\
 2.1 &  $24\times 48$ &  0.1695 &  0.381(23) &  0.637(67) &  2.289(11) &  314 \\
 2.2 &  $24\times 48$ &  0.1600 &  1.048(12) &  - &  - &  187 \\
 2.2 &  $24\times 48$ &  0.1640 &  0.530(13) &  0.593(44) &  3.065(19) &  331 \\
 2.2 &  $24\times 48$ &  0.1650 &  0.315(20) &  0.410(28) &  3.011(16) &  320 \\
 2.2 &  $24\times 48$ &  0.1652 &  0.244(20) &  0.316(34) &  3.348(27) &  333 \\
 2.2 &  $32\times 64$ &  0.1652 &  - &  - &  - &  128 \\
\bottomrule
\end{tabular}

%% file: tablepuremixed.tex
\begin{tabular}{llrrlllllllllll}
\toprule
$\beta$ & $N_s\times N_t$ &  $\kappa_A$ &  $\kappa_F$ & $am_{PCAC,F}$ & $am_{PS,F}$ &  $af_{PS,F}$ &  $am_{V,F}$ & $am_{PCAC,A}$ & $am_{PS,A}$ &  $af_{PS,A}$ &   $am_{V,A}$ &  $am_{gg}$ &    $w_0/a$ & $N_{conf}$ \\
\midrule
 2.1 &  $24\times 48$ &  0.158 &  0.134 &  0.12160(11) &  0.4775(37) &  0.04330(34) &  0.4955(36) &  0.14217(21) &  0.8047(52) &  0.1673(13) &  0.8661(68) &  0.721(21) &  - &  850 \\
 2.1 &  $24\times 48$ &  0.160 &  0.134 &  0.110448(88) &  0.4399(45) &  0.04074(32) &  0.4621(34) &  0.08657(18) &  0.5885(59) &  0.1306(15) &  0.6453(70) &  0.535(25) &  4.369(61) &  480 \\
 2.1 &  $24\times 48$ &  0.162 &  0.134 &  0.09691(17) &  0.3985(77) &  0.03634(44) &  0.4206(61) &  0.02611(19) &  0.270(12) &  0.0697(18) &  0.312(17) &  0.261(24) &  4.583(56) &  304 \\
 2.1 &  $24\times 48$ &  0.100 &  0.135 &  0.18668(21) &  0.8221(73) &  0.0908(13) &  0.8848(79) &  1.23247(73) &  2.9703(35) &  0.2611(11) &  2.9964(38) &  2.54(18) &  - &  379 \\
 2.1 &  $24\times 48$ &  0.120 &  0.135 &  0.17354(29) &  0.760(11) &  0.0790(14) &  0.8150(96) &  0.89868(85) &  2.4428(58) &  0.3192(23) &  2.4955(62) &  2.156(60) &  - &  203 \\
 2.1 &  $24\times 48$ &  0.140 &  0.135 &  0.14713(28) &  0.634(11) &  0.0659(15) &  0.6695(78) &  0.53590(76) &  1.815(11) &  0.3350(63) &  1.898(15) &  1.52(11) &  - &  213 \\
 2.1 &  $24\times 48$ &  0.150 &  0.135 &  0.12226(11) &  0.5238(56) &  0.05152(59) &  0.5519(44) &  0.33084(29) &  1.3672(67) &  0.2854(40) &  1.4529(97) &  1.180(29) &  - &  473 \\
 2.1 &  $24\times 48$ &  0.155 &  0.135 &  0.10484(13) &  0.4545(75) &  0.04509(57) &  0.4803(44) &  0.21306(31) &  1.0385(75) &  0.2109(27) &  1.10841(10) &  0.867(25) &  - &  412 \\
 2.1 &  $24\times 48$ &  0.158 &  0.135 &  0.09178(10) &  0.4093(43) &  0.04156(35) &  0.4396(32) &  0.13628(17) &  0.7805(50) &  0.1624(16) &  0.8418(63) &  0.682(16) &  3.274(20) &  845 \\
 2.1 &  $24\times 48$ &  0.160 &  0.135 &  0.08039(18) &  0.3521(67) &  0.03684(41) &  0.3960(52) &  0.08063(26) &  0.5578(81) &  0.1220(17) &  0.6183(91) &  0.469(14) &  3.711(24) &  398 \\
 2.1 &  $24\times 48$ &  0.161 &  0.135 &  0.07379(13) &  0.3260(76) &  0.03706(45) &  0.3527(63) &  0.05136(22) &  0.4240(94) &  0.1009(15) &  0.482(11) &  0.426(17) &  4.058(20) &  302 \\
 2.1 &  $24\times 48$ &  0.162 &  0.135 &  0.06582(10) &  0.3319(69) &  0.03234(41) &  0.3625(50) &  0.01926(18) &  0.230(11) &  0.0665(14) &  0.288(14) &  0.214(19) &  4.754(27) &  318 \\
 2.1 &  $24\times 48$ &  0.158 &  0.136 &  0.06004(12) &  0.2983(56) &  0.03457(35) &  0.3399(46) &  0.12980(20) &  0.7495(60) &  0.1539(16) &  0.8062(65) &  0.668(22) &  3.879(55) &  754 \\
 2.1 &  $32\times 48$ &  0.158 &  0.136 &  0.06147(19) &  0.3239(72) &  0.03756(63) &  0.3754(48) &  - &  - &  - &  - &  - &  - &  849 \\
 2.1 &  $24\times 48$ &  0.160 &  0.136 &  0.04884(13) &  0.2515(71) &  0.03403(51) &  0.3039(55) &  0.07356(20) &  0.5186(72) &  0.1144(15) &  0.5734(73) &  0.411(15) &  3.987(21) &  572 \\
 2.1 &  $24\times 48$ &  0.161 &  0.136 &  0.04217(10) &  0.2514(68) &  0.02938(46) &  0.3080(39) &  0.04351(21) &  0.3773(89) &  0.0897(11) &  0.438(10) &  0.324(17) &  4.368(39) &  328 \\
 2.1 &  $24\times 48$ &  0.162 &  0.136 &  0.03426(14) &  0.1988(86) &  0.02939(49) &  0.2479(54) &  0.01242(25) &  0.178(13) &  0.0604(20) &  0.265(15) &  0.178(14) &  4.615(35) &  438 \\
 2.1 &  $32\times 48$ &  0.162 &  0.136 &  0.03467(11) &  0.211(11) &  0.02564(53) &  0.2666(43) &  0.01314(19) &  0.1991(85) &  0.0569(13) &  0.2805(98) &  0.217(16) &  - &  447 \\
 2.1 &  $24\times 48$ &  0.160 &  0.100 &  0.9927(14) &  2.464(11) &  0.1151(22) &  2.478(11) &  0.23826(55) &  1.214(13) &  0.3779(84) &  1.422(25) &  0.97(27) &  - &  150 \\
 2.1 &  $24\times 48$ &  0.160 &  0.120 &  0.49525(19) &  1.4090(55) &  0.07159(80) &  1.4193(60) &  0.1633(12) &  0.951(30) &  0.243(12) &  1.050(43) &  0.955(65) &  - &  303 \\
 2.1 &  $24\times 48$ &  0.160 &  0.132 &  0.16970(14) &  0.6083(36) &  0.04694(34) &  0.6202(34) &  0.09917(23) &  0.6484(77) &  0.1411(23) &  0.7111(90) &  0.706(17) &  4.18(12) &  445 \\
 2.1 &  $24\times 48$ &  0.160 &  0.133 &  0.14080(11) &  0.5230(50) &  0.04516(38) &  0.5452(37) &  0.09281(20) &  0.6181(76) &  0.1389(17) &  0.6813(85) &  0.543(19) &  3.855(43) &  393 \\
 2.2 &  $24\times 48$ &  0.157 &  0.132 &  0.139759(85) &  0.5019(48) &  0.04121(40) &  0.5215(40) &  0.09512(16) &  0.5802(72) &  0.1137(17) &  0.6322(83) &  0.435(28) &  4.127(43) &  347 \\
 2.2 &  $24\times 48$ &  0.158 &  0.132 &  0.135202(62) &  0.4901(33) &  0.03879(31) &  0.5020(29) &  0.06658(17) &  0.4582(94) &  0.0918(16) &  0.4893(99) &  0.439(52) &  4.921(15) &  600 \\
 2.2 &  $24\times 48$ &  0.158 &  0.133 &  0.105987(74) &  0.4141(41) &  0.03564(28) &  0.4355(32) &  0.06197(17) &  0.4340(59) &  0.08687(96) &  0.4787(59) &  0.364(24) &  5.299(47) &  561 \\
 2.2 &  $24\times 48$ &  0.158 &  0.134 &  0.076362(91) &  0.3393(73) &  0.03281(36) &  0.3854(45) &  0.05777(20) &  0.4134(79) &  0.0880(15) &  0.4688(80) &  0.363(19) &  4.639(25) &  336 \\
 2.2 &  $24\times 48$ &  0.159 &  0.134 &  0.07203(10) &  0.3363(68) &  0.03281(40) &  0.3784(53) &  0.02901(24) &  0.2883(83) &  0.0734(12) &  0.362(10) &  0.309(23) &  4.034(31) &  362 \\
 2.2 &  $24\times 48$ &  0.158 &  0.135 &  0.045494(91) &  0.2234(57) &  0.03000(40) &  0.2750(45) &  0.05234(15) &  0.3842(67) &  0.0798(10) &  0.4304(70) &  0.317(25) &  4.886(39) &  536 \\
 2.2 &  $24\times 48$ &  0.159 &  0.135 &  0.039821(90) &  0.2473(82) &  0.02603(43) &  0.2680(61) &  0.02204(23) &  0.2102(87) &  0.05549(85) &  0.264(12) &  0.272(33) &  5.561(50) &  329 \\
 2.2 &  $32\times 64$ &  0.159 &  0.135 &  0.04338(17) &  0.244(11) &  0.03160(52) &  0.3125(65) &  0.02592(22) &  0.2725(99) &  0.0722(19) &  0.364(10) &  0.286(13) &  - &  338 \\
\bottomrule
\end{tabular}